\documentclass{emulateapj-rtx4}

\usepackage{color}
\usepackage{ulem}

\def\ion#1#2{#1$\;${\sc\@roman{#2}}\relax}
\def\lesssim{\mathrel{\hbox{\rlap{\hbox{\lower4pt\hbox{$\sim$}}}\hbox{$<$}}}}
\def\gtrsim{\mathrel{\hbox{\rlap{\hbox{\lower4pt\hbox{$\sim$}}}\hbox{$>$}}}}

\newcommand{\xHI}{x_\mathrm{HI}}

\slugcomment{accepted for publication in ApJ; v04/14/2016}

\shorttitle{Ly$\alpha$ Fraction of UV Very Bright Galaxies at $z=7$}
\shortauthors{Furusawa et al.}

\begin{document}

\title{A New Constraint on the Ly$\alpha$ Fraction of 
UV Very Bright Galaxies at Redshift 7}

\author{
Hisanori Furusawa\altaffilmark{1}, 
Nobunari Kashikawa\altaffilmark{1,2}, 
Masakazu A. R. Kobayashi\altaffilmark{3}, 
James S. Dunlop\altaffilmark{4}, 
Kazuhiro Shimasaku\altaffilmark{5}, 
Tadafumi Takata\altaffilmark{1,2}, 
Kazuhiro Sekiguchi\altaffilmark{1,2}, 
Yoshiaki Naito\altaffilmark{6}, 
Junko Furusawa\altaffilmark{1}, 
Masami Ouchi\altaffilmark{6,7},
Fumiaki Nakata\altaffilmark{8},
Naoki Yasuda\altaffilmark{7},
Yuki Okura\altaffilmark{9,10},
Yoshiaki Taniguchi\altaffilmark{3},
Toru Yamada\altaffilmark{11}, 
Masaru Kajisawa\altaffilmark{3}, 
Johan P. U. Fynbo\altaffilmark{12},
and 
Olivier Le F\`evre\altaffilmark{13}
}

\altaffiltext{1}{National Astronomical Observatory of Japan, 2-21-1 Osawa,
Mitaka, Tokyo 181-8588, Japan; furusawa.hisanori\_at\_nao.ac.jp}
\altaffiltext{2}{Department of Astronomy, School of Science, SOKENDAI
(The Graduate University for Advanced Studies), Mitaka, Tokyo 181-8588,
Japan}
\altaffiltext{3}{Research Center for Space and Cosmic Evolution, Ehime
University, 2-5 Bunkyo-cho, Matsuyama, Ehime 790-8577, Japan}
\altaffiltext{4}{Institute for Astronomy, University of Edinburgh, Royal
Observatory, Blackford Hill, Edinburgh EH9 3HJ, UK}
\altaffiltext{5}{Department of Astronomy, School of Science, University
of Tokyo, 7-3-1 Hongo, Bunkyo-ku, Tokyo 113-0033, Japan}
\altaffiltext{6}{Institute for Cosmic Ray Research, The University of Tokyo, 5-1-5 Kashiwanoha, Kashiwa, Chiba 277-8582, Japan}
\altaffiltext{7}{Kavli Institute for the Physics and Mathematics of the
Universe (Kavli IPMU, WPI), The University of Tokyo, 5-1-5 Kashiwanoha, Kashiwa,
Chiba 277-8583, Japan}
\altaffiltext{8}{Subaru Telescope, National Astronomical Observatory of
Japan, 650 North A'ohoku Place, Hilo, HI 96720, U.S.A.}
\altaffiltext{9}{RIKEN, 2-1 Hirosawa, Wako, Saitama 351-0198, Japan}
\altaffiltext{10}{RIKEN-BNL Research Center, Department of Physics, Brookhaven National Laboratory, Bldg. 510, Upton, NY, 11792, U.S.A}
\altaffiltext{11}{Astronomical Institute, Tohoku University, 6-3 Aramaki,
Aoba-ku, Sendai 980-8578, Japan}
\altaffiltext{12}{1Dark Cosmology Centre, Niels Bohr Institute,
University of Copenhagen, Juliane Maries Vej 30, DK-2100 Copenhagen,
Denmark}
\altaffiltext{13}{Aix-Marseille Universit\'e, CNRS, LAM (Laboratoire d'Astrophysique de Marseille) UMR 7326, 13388 Marseille, France}

\begin{abstract}
We study the extent to which very bright ($-23.0<M_{UV}<-21.75$) 
 Lyman-break selected galaxies at redshifts $z\simeq 7$ display
 detectable Ly$\alpha$ emission. To explore this issue, we have obtained
 follow-up optical spectroscopy of 9 $z\simeq 7$ galaxies from a parent
 sample of 24 $z\simeq 7$ galaxy candidates selected from the 1.65
 deg$^2$ COSMOS-UltraVISTA and SXDS-UDS survey fields using the latest
 near-infrared public survey data, and new ultra-deep Subaru $z'$-band
 imaging (which we also present and describe in this paper). 
Our spectroscopy has yielded only one possible detection of
 Ly$\alpha$ at $z=7.168$ with a rest-frame equivalent width
 $\mathrm{EW}_0=3.7^{+1.7}_{-1.1}\,{\rm \AA}$.
The relative weakness of this line, combined with our failure to 
detect Ly$\alpha$ emission from the other spectroscopic targets allows
 us to place a new upper limit on the prevalence of strong Ly$\alpha$
 emission at these redshifts. For conservative calculation and to
 facilitate comparison with previous studies at lower redshifts, we
 derive a 1-$\sigma$ upper limit on the fraction of UV-bright
 galaxies at $z\simeq 7$ that display $\mathrm{EW}_0>50\,{\rm \AA}$,
 which we estimate to be $<0.23$.
This result may indicate a weak trend where the fraction of strong 
Ly$\alpha$ emitters ceases to rise, and possibly falls between $z\simeq
 6$ and $z\simeq 7$. Our results also leave open the possibility that strong
 Ly$\alpha$ may still be more prevalent in the brightest galaxies in the
 reionization era than their fainter counterparts.
A larger spectroscopic sample of galaxies is required to derive 
a more reliable constraint on the neutral hydrogen 
fraction at $z\sim 7$ based on the Ly$\alpha$ fraction in the 
bright galaxies.

\end{abstract}


\section{Introduction}\label{sec:intro}

Understanding the epoch of cosmological reionization has been a key 
challenge in modern observational cosmology.
The Gunn-Peterson test with spectra of high-redshift (high-$z$) quasars
is an observational tool for detecting the reionization epoch, 
and suggests that the intergalactic medium (IGM) was reionized at
$z>6$ \citep{fan06}. 
There is also observational evidence that neutral hydrogen still
remains in the IGM of the universe at $z\sim 6$, e.g., studies with
quasars \citep{becker15} and with gamma-ray bursts (GRBs) \citep{totani14, hartoog15}.  
Recent analysis of the cosmic microwave background (CMB) indicated the 
redshift of the cosmic reionization of $\sim 8.8$ by Planck \citep{planck15}.

Galaxies associated with Ly$\alpha$ emission, i.e., 
Lyman-$\alpha$ emitters (LAEs) are also valuable probes to 
infer the neutral hydrogen (H {\sc i}) fraction $\xHI$ of the IGM. 
\citet{mcquinn07} suggested the effectiveness of this concept from a 
theoretical viewpoint. 
Observational studies indicated that the luminosity functions (LFs) 
of the Lyman-break galaxies (LBGs) show a monotonic decline 
in number density at $z=3$ to $z=7$ \citep{mclure13, bowler12, bowler14, bowler15, bouwens15}. 
In contrast, the LFs of LAEs show little change at $z=3-6$, 
implying an increasing fraction of LAEs relative to LBGs with 
increasing redshift; in turn, they show a significant decline at $z>6$
in the number density, 
e.g., \citet{konno14} who presented a clear decrease in the Ly$\alpha$ LF 
at $z=7.3$ from the $z=6.6$.
This rapid decline in the Ly$\alpha$ LF can be attributed to a 
change in $\xHI$ at $z>6$. The neutral hydrogen in the IGM 
could decrease observed Ly$\alpha$ photons from star-forming galaxies 
by resonant scattering\citep{kobayashi10}.
\citet{ouchi10} suggested an increase in the H {\sc i} fraction from 
$z=5.7$ to $z=6.5$ and the estimated $\xHI$ is 0.22 at $z=6.5$.
They suggested possible cosmic variance in the number density 
of LAEs across a $\sim 1$ sq.degree area by a factor of 2 to 10 
depending on their NB921 magnitudes.
\citet{kashik11} inferred $\xHI =0.38$ at $z=6.5$ from a different data set, 
suggesting existence of the cosmic variance in $\xHI$ with a possible patchy 
reionization process in the universe.

A caveat in investigating reionization with the LFs of galaxies 
is that the LFs are likely to be influenced by galaxy evolution. 
In contrast, the Ly$\alpha$ fraction, i.e., the fraction of galaxies 
exhibiting strong Ly$\alpha$ emission of the total LBGs, has 
an advantage in that it should be less affected by galaxy evolution, 
especially in terms of the number density, than the Ly$\alpha$ LFs, 
which require comparison with the UV LFs \citep{stark10}.
In addition, star formation activity and dust/metal content of galaxies 
 which contribute to the equivalent width (EW) of Ly$\alpha$, do not
 change rapidly even at $z=7-8$ compared to those at lower redshifts
 \citep{dunlop13}.

The Ly$\alpha$ fraction at each redshift has been studied intensively to 
yield a complementary information to the cosmic reionization
process \citep{stark10, stark11, pentericci11, ono12,
curtis-lake12, treu13, tilvi14, caruana14, cassata15}. 
The studies by \citet{stark10} and \citet{ono12} suggested 
that the Ly$\alpha$ fraction of LBGs drops significantly from $z=6$ to
$z=7$ in contrast to its monotonic increase at $z=4-6$. Their faint
sample showed a more rapid decline than the bright sample.
\citet{tilvi14} also studied the Ly$\alpha$ fraction at $z\gtrsim 7$ 
and discussed the decline in Ly$\alpha$ fraction at $z\gtrsim 7.5$ 
with the possibility of witnessing the ongoing cosmic reionization process 
at $z\sim 7-8$. 
A recent study with Hubble Space Telescope (HST) slitless spectroscopy
targeting gravitationally lensed clusters of galaxies (GLASS) also
indicated that the number of LAEs at $z\sim 7$ with respect to that of the 
LBGs is consistent with a lower probability of Ly$\alpha$ emission 
at $z\gtrsim 7$ than at $z\sim 6$ \citep{schmidt16}.

While the relative number of LAEs is an effective tracer of the cosmic
reionization state, a fair application would require consideration of
several factors that may affect the observable Ly$\alpha$ emission from 
high-$z$ galaxies.
The strength of Ly$\alpha$ emitted from galaxies may be affected by 
a combination of physical properties of galaxies and the transmission
 of the IGM along the lines of sight.

\citet{stark10} discussed the importance of the coupled effects by 
internal dust extinction and the geometry and kinematics of H {\sc i} 
gas surrounding a galaxy, to determine the net Ly$\alpha$ 
photons that can escape from the galaxy. 
LBGs with low luminosities tend to have low metallicity or 
dust extinction, leading to a bluer UV color \citep{bouwens14,
bowler15} and likely association of strong Ly$\alpha$ emission. 
This may introduce a possible luminosity dependence of the Ly$\alpha$
fraction, as suggested in previous studies \citep{stark11, ono12}.
\citet{steidel10} discussed a similar trend in the stellar mass
dependence on the strong Ly$\alpha$ emission.

Observational studies suggested that LBGs with high UV luminosities tend
to show a deficit in strong Ly$\alpha$ emission \citep{ando06, shimasaku06}.
Based on spectroscopic observations of LBGs, \citet{stark11} reported 
that the Ly$\alpha$ fraction of LBGs with low UV luminosities is higher
than that of those with high luminosities at $z=3-6$.
In addition, the Ly$\alpha$ fraction ($\mathrm{EW}>25\,{\rm \AA}$) rises as redshift
increases over $z=4-6$.
\citet{cassata15} supported this trend at $z<5$ based on a spectroscopic
survey of galaxies in the VIMOS-VLT Deep Survey (VVDS) sample.
\citet{curtis-lake12} reported the same increase in Ly$\alpha$ fraction
($\mathrm{EW}>25\,{\rm \AA}$) from $z=5$ to $6$ at bright magnitudes ($L>2L^*$).
Recent spectroscopic studies have also reported detection of Ly$\alpha$ 
in a few sources in a similar luminosity range \citep{oesch15,
roberts-borsani15, zitrin15}. 

The redshift above $\sim 7$ is an important epoch to understand the
progression of reionization. In particular, it is of interest to 
investigate even brighter objects ($M_{UV}<-21.75$), i.e., in
environments in very massive dark matter halos.
Such massive halos are thought to evolve and form galaxies
with strong clustering in an earlier epoch, and so are likely to host larger 
amounts of ionizing sources than less massive halos. 
Therefore, it is expected that reionization would progress earlier in 
the environments around the massive dark matter halos than in other regions.
However, despite intensive observational studies of the Ly$\alpha$
fraction at $z>6$, only small numbers of the LBGs at $z\gtrsim 7$ are
available. 
Primarily due to limitations in the surveyed area and as brighter
galaxies are rarer, very little is known about the LBGs with very bright
magnitudes. 
Previous studies on the Ly$\alpha$ fraction were conducted only
at UV magnitudes fainter than $-21.75$. 
Thus, new deep and wide imaging and spectroscopic observations 
are indispensable to extend our knowledge of the bright magnitude regime.
Focusing on such UV-bright LBGs is also an effective approach 
that facilitates follow-up spectroscopy and allows determination of a
reliable Ly$\alpha$ fraction at high redshifts.

We have undertaken a new survey program to cover an unprecedentedly
 wide area by a deep imaging survey to detect UV-bright LBGs, 
followed by spectroscopic observation of some of the high-$z$ candidate galaxies. 
In this study, with the new spectroscopic sample from our survey
program, 
we examine the Ly$\alpha$ fraction of LBGs at $z=7$ for very bright
magnitudes. We also discuss the cosmic reionization at redshift 7,
providing a new constraint on the cosmic neutral H {\sc i} fraction.
In Section~2, we summarize the observations and data reduction
in our $z'$-band imaging survey program, and explain the method of 
target selection for LBG candidates at $z=7$. 
Section~3 describes the follow-up spectroscopy and 
properties of the spectroscopic sample. 
The Ly$\alpha$ EW of our UV-bright galaxy sample is described in
Section~4. We discuss the Ly$\alpha$ fraction of LBGs in Section~5, and 
its implications and interpretation regarding the state of 
the cosmic reionization in Section~6. Finally, conclusions are presented
in Section~7.
Throughout the paper, we use the AB magnitude system \citep{oke83}, 
and assume a flat universe with $(\Omega_m, \Omega_\Lambda , h)=(0.3, 0.7, 0.7)$.

\section{Imaging Observations, Data, and Target Selection}
This paper aims to study the Ly$\alpha$ fraction based on LBG candidates 
at $z=7$ in the two common fields, COSMOS/UltraVISTA and SXDS/UKIDSS-UDS
(hereafter, UltraVISTA and UDS, respectively). 
These two survey fields represent a unique and powerful combination for
studies on high-$z$ galaxies, where extensive multi-wavelength surveys at
various facilities have been performed.
In particular, the two ground-based deep near-infrared (NIR) surveys
UltraVISTA \citep{mccracken12} and UKIDSS-UDS \citep{lawrence07} 
provide an unprecedentedly wide and deep data set over an area of 1.65 
sq.degree, making the two fields the most suitable for searches of bright LBGs
at $z=7$. 

Initially, however, only relatively shallow optical data sets 
were available in these fields, especially in the $z'$ band, which is 
crucial for detecting the dropout feature of the Lyman-break of 
the $z=7$ galaxies, e.g., COSMOS imaging \citep{capak07} by the Subaru Telescope
covering the UltraVISTA field in the $B, g', V, r', i'$ and $z'$ bands, with
$z'=25.1$ (5 $\sigma$, 3-arcsec aperture), 
and the Subaru/XMM-Newton Deep Survey (SXDS) \citep{furusawa08}
located at the UDS field in the $B, V, R_c, i'$, and $z'$ bands to
a depth of $z'\sim 26$ (5 $\sigma$, 2-arcsec aperture). 
This situation motivated us to undertake a new observational campaign 
of ultra-deep imaging in the $z'$ band in the two fields.

We use a combined sample of the LBGs presented previously by 
\citet{bowler12, bowler14}, and a small addition to this work in the UDS
field as described below.
In the following sections, we first present an overview of the entire 
$z'$-band imaging campaign and data reduction. Then, construction of
the sample of LBGs in each field is explained.

\subsection{Suprime-Cam $z'$-band Data}
\subsubsection{Observations}

\begin{figure}
\epsscale{1.}
\plotone{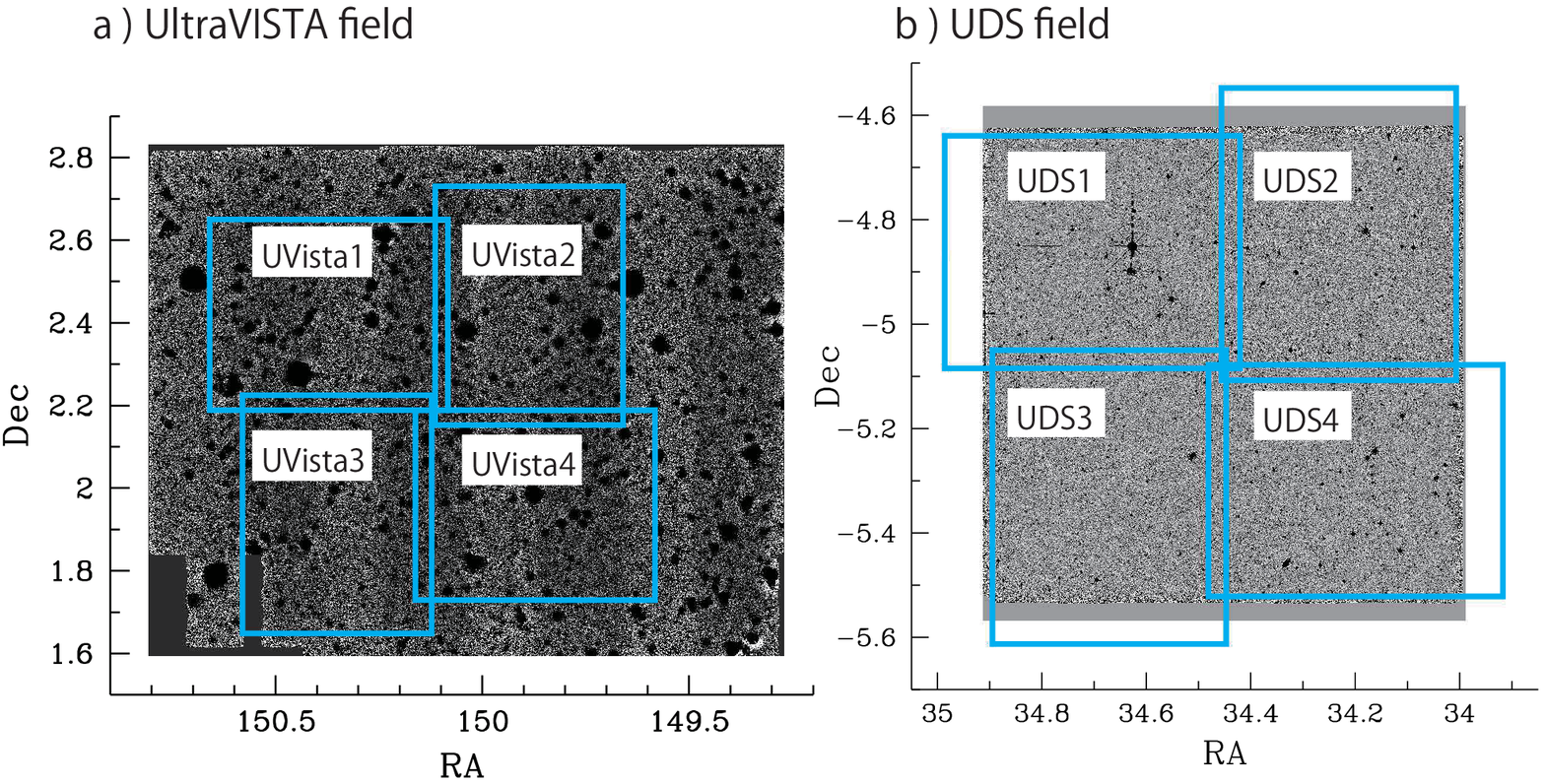}
\caption{The arrangement of the camera pointings with Suprime-Cam on
 the two survey fields. The left panel (a) shows the UltraVISTA field
 image overlaid with four fiducial pointings of Suprime-Cam (cyan
 rectangles) labeled with each field name, 
while the right panel (b) is the same but for the UDS field. The
 coordinates are shown in units of degrees. The base images are
 derived from a) the UltraVISTA-DR2 $J$-band image and b) the UDS-DR10 $J$-band image.
\label{fig:pointing}}
\end{figure}

The deep imaging observations in the $z'$ band were conducted 
with Suprime-Cam \citep{miyazaki02} using the 8.2 m Subaru Telescope, 
which uses fully depleted CCDs that were introduced in 2008 August. 
Our survey plan was designed to cover the two target fields 
UltraVISTA and UDS ($\sim 2$ sq.degree) with eight pointings of
Suprime-Cam, the field of view of which is $34\times 27$ sq.arcmin. 
Figure~\ref{fig:pointing} summarizes this pointing strategy.
Each pointing was planned to be integrated by 20 hours, divided into 
shorter exposures.
The observing times were awarded as an open-use Subaru intensive program 
(PI: Furusawa; S08B-051). 
Originally, 20 nights were allocated over three semesters from 2008 October to 
2009 November. To compensate for low completion rate due to 
poor weather conditions, etc., another 16 nights were added 
by a regular open-use program and the observatory's 
discretionary time. Final exposure times in each pointing, which are merged into 
the stacked images are 19.6, 18.3, 18.7, and 20.8 hours for the UltraVISTA
field, and 11.7, 16.2, 11.7, and 15.5 hours for the UDS field. 
The $z'$-band data obtained in this campaign are summarized in
Table~\ref{tab:zdata}.

\subsubsection{Data Reduction}

Data reduction of the $z'$-band data 
performed to generate mosaic-stacked images from raw CCD images, 
following the procedure described in \citet{furusawa08}.
The SDFRED2 software \citep{yagi02, ouchi04} provided by the observatory
was used. 
Small fluctuations due to cross-talk between amplifiers of the CCD 
were corrected \citep{yagi12} for the UDS field data.
The full width half-maximums (FWHMs) of the point spread
function (PSF) of the stacked images before any convolution or
transformation are 0.76, 0.84, 0.70, and 0.82 arcsec for the
UltraVISTA field, and 0.75, 0.81, 0.78, and 0.81 arcsec for the UDS
field.

The celestial coordinates of the stacked images were calibrated with 
the USNO-B1.0 catalog \citep{monet03}, 
using SExtractor and SCAMP\footnote{\tt http://www.astromatic.net} .
The resultant residuals of the determined coordinates 
from the USNO-B1.0 catalog coordinates are on the order of 0.5 to 0.8 arcsec rms. 
This accuracy is good enough to perform image warping in
Section~\ref{sec:sample_selection_uds}.

Magnitude zero points of the stacked images 
were determined with a standard star GD71 ($z'=14.03$).
The estimated depths in the $z'$ band are 26.6, 26.5, 26.5, 
and 26.8 (5 $\sigma$, 2-arcsec aperture) for the UltraVISTA field, and
26.3, 26.6, 26,6, and 26.6 for the UDS field.
This data set is the deepest ever achieved in the $z'$ band 
covering the entire two deep fields ($\sim 2$ sq.degree).

\subsection{Sample Selection in UltraVISTA Field}

Sample selection of $z=7$ galaxies in the UltraVISTA field (RA
$10^h00^m28.^s00$, Dec. $+02^\circ 12^{'}30^{''}$, J2000) was performed 
by \citet{bowler14}, in an update to their initial work \citep{bowler12}. 
In the field, the UltraVISTA DR2 provides NIR photometry in the $Y,
J, H$, and $K_s$ bands ($J=25.3$ in ultradeep strips; 5 $\sigma$,
1.8-arcsec aperture) over a field of 0.91 sq.degree 
e.g., Figure~1 in \citet{bowler14}. 
They combined the UltraVISTA-DR2 data with the multi-waveband data
including the CFHTLS data in the $u^{*}, g, r, i, z'$ bands ($z=25.2$;
5$\sigma$, 2-arcsec aperture), and the final $z'$-band data set 
(Table~\ref{tab:zdata}) provided by our imaging campaign. 

The high-$z$ candidates were selected based on a photometric 
redshift ($z_\mathrm{phot}$) analysis applied to the sources detected 
in the $J$ or $Y+J$ band under conditions with no detection in the
$i$ or bluer bands. 
The resultant $z=7$ LBG sample in the UltraVISTA field contained 19
candidates with $z_\mathrm{phot}>6.5$, including 13 sources 
at $z_\mathrm{phot}>6.75$.

\subsection{Sample Selection in UDS Field}\label{sec:sample_selection_uds}

The UDS-DR10 data (Omar Almaini, private communication) covers an area of 0.74
sq.degree centered on the UDS field (RA $02^h17^m48^s$, Dec. 
$-05^\circ 05^{'}57^{''}$, J2000) to depths of $J=25.5, H=24.9$, and
$K=25.1$ (5 $\sigma$, 2-arcsec aperture). 
\citet{bowler14} performed selection of $z=7$ LBGs in the UDS 
field with photometric redshift analysis using the multi-waveband data
set including the NIR UDS ($J, H$, and $K$) data and the optical
SXDS ($B, V, R_c, i'$, and $z'$) data. 
The data set also included the $z'$-band data in our imaging 
campaign. However, the $z'$-band data were of interim status, 
which consisted of 11.1, 16.2, 11.2, and 15.5 hours slightly shallower 
in the UDS1 and UDS3 than the final data set listed in
Table~\ref{tab:zdata}. Additional $z'$-band data have been obtained
since their study.

Their LBG sample in the UDS field included only two candidates with
$z_\mathrm{phot}>6.5$ in contrast to the UltraVISTA field. 
This relatively small number of high-$z$ candidates in the UDS field may 
have been due to their requirement of significant 
detection in the $Y$ band to avoid contamination by severe
cross-talk from the UKIRT/WFCAM, which is only applied to the UDS field.
To add as many $z=7$ candidates to their LBG sample in this
field as possible, in this study we used the combined data of the UDS-DR10, 
the SXDS, and the updated $z'$-band data set with the final depths 
(Table~\ref{tab:zdata}). Moreover, we applied a different
sample selection to the data from that adopted in \citet{bowler14}.
Our additional sample selection procedure is described below.

\subsubsection{Additional Sample Selection in UDS Field}\label{sec:uds_selection}

We transformed the $z'$-band stacked images in the UDS field onto the
same pixel coordinates as those of the $J$-band image.
The images in the $z'$ to $K$ bands were convolved with a single Gaussian
smoothing kernel to have the same FWHMs (0.83 arcsec) of the PSF as 
the $J$-band images.

On the $z', J, H, K$ images, we ran SExtractor (ver. 2.19.5) 
in double-image mode, with detection on the $J$-band image, 
to generate $J$-band detected multi-waveband catalogs ($J\leq 25.5$; 5
$\sigma$, 2-arcsec aperture), separately for each of the four pointings.
We also removed sources located at pixels that were affected by 
low signal-to-noise ratio (S/N) around the image edges, strong blooming, 
and strong halos of bright stars.

We applied the following condition to the magnitude-limited sample of
 galaxies:
 (1) $z'-J>2.5$, (2) $J-K>-0.5$, and (3) complete dropout in all 
of the $B, V, R_c$ and $i'$ bands. 
The first condition involves selection of the Lyman-break, which
 together with the second condition minimizes contamination by galactic
 LT-dwarf stars. 
The third condition is to reduce contamination by lower-redshift 
galaxies.  
The resultant intermediate catalogs that satisfied the above
 conditions included 659, 757, 725, and 700 sources in each 
pointing, respectively.

To remove false sources due to severe cross-talk on the $J, H, K$
images, we visually inspected every candidate source in the
 intermediate catalogs by eye, instead of requiring 
detection in the $Y$ band.
The false signals due to cross-talk appeared at positions of 191 pixels 
and/or its multiples from bright root signals in either the x- or y-axis, 
in each of the $J, H, K$ bands. 
These locations of cross-talk signals are inherited from the characteristics 
of the read-out electronics of UKIRT/WFCAM.
If the candidate is an artifact of cross-talk, there must be a root 
signal that causes a series of cross-talk located at 
one of the positions of the multiples of 191 pixels from the candidate.
Importantly, a series of false signals 
decays monotonically with distance from the root signal, 
and also does not go beyond the quadrant area where the root signal is
 located. Therefore, for every candidate source, we inspected 
24 locations with distances of 191 and 282 pixels from the
 candidate on the x- and y-axes and grid points according to their 
combinations. In cases where there is any possible false signal at these
 locations, we excluded the candidate source as it is likely to be a
 false signal.

\begin{figure}
\epsscale{1.}
\plotone{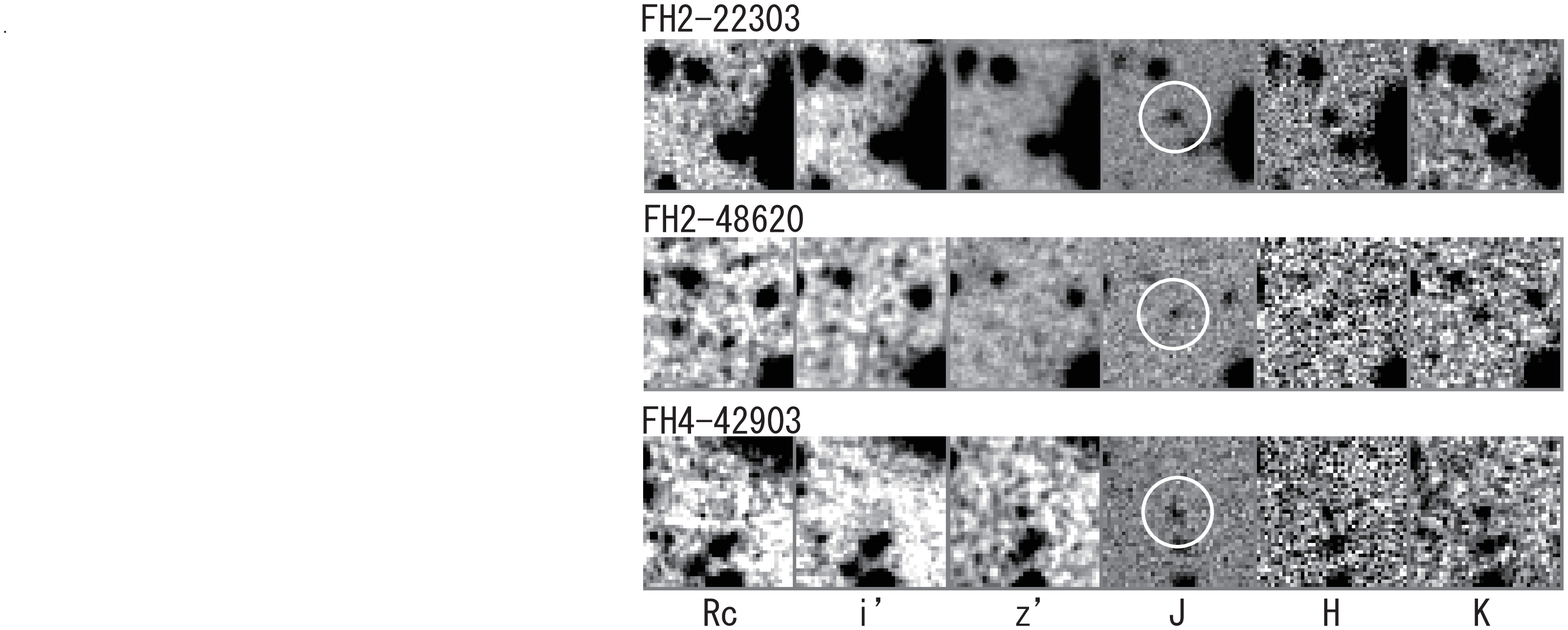}
\caption{Postage stamp images of the three new high-$z$ candidate sources 
in the UDS field. For each source, the dumped images of the source are
 shown in the $R_c, i', z', J, H$, and $K$ bands from the left to the
 right side. The image size in each band is 10.7 arcsec on a side, and 
 white circle with a diameter of 5 arcsec is overplotted on the $J$ band
 images to show the source position.\label{fig:stamps}}
\end{figure}

\begin{figure}
\epsscale{0.5}
\plotone{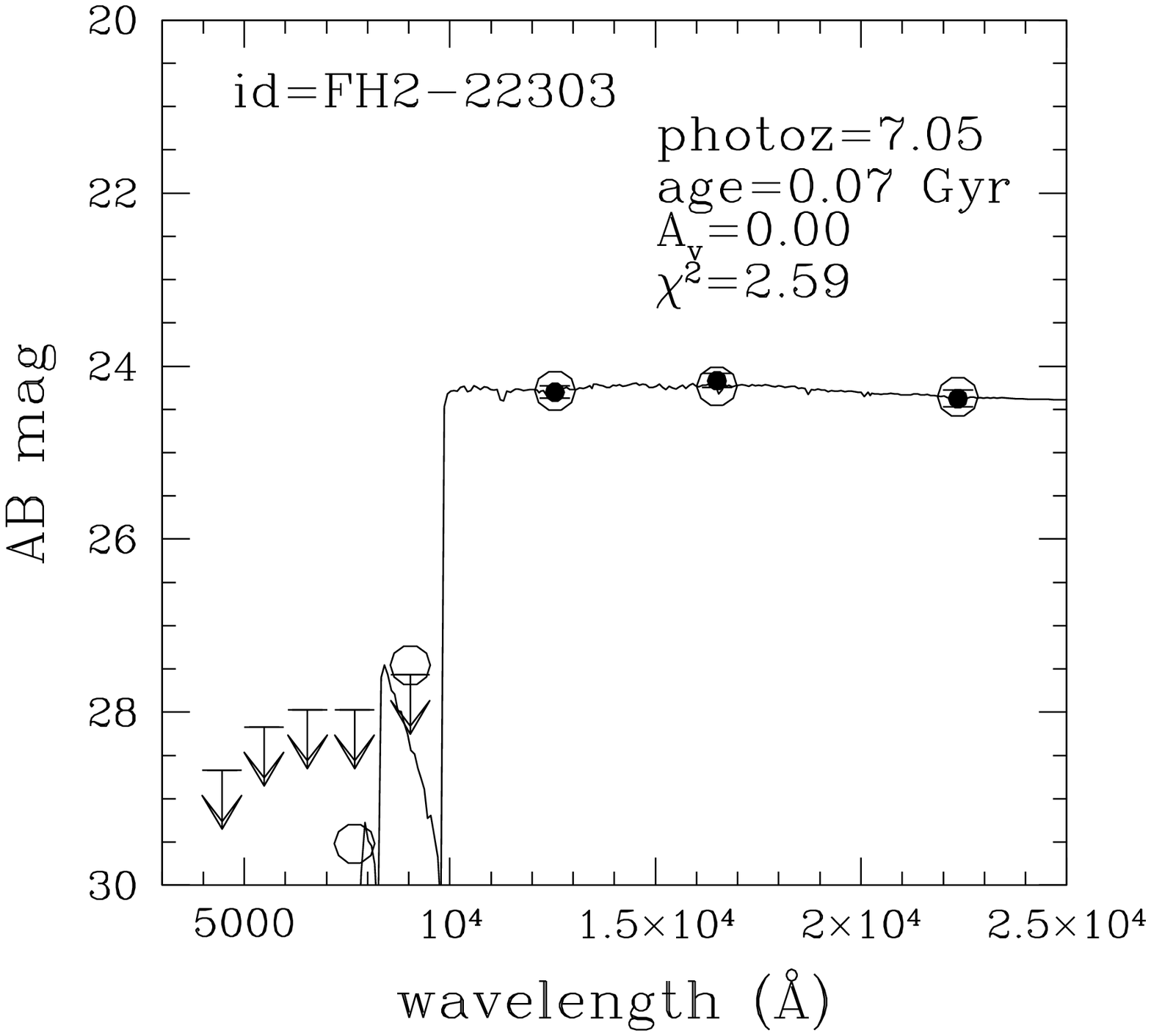}\\
\plotone{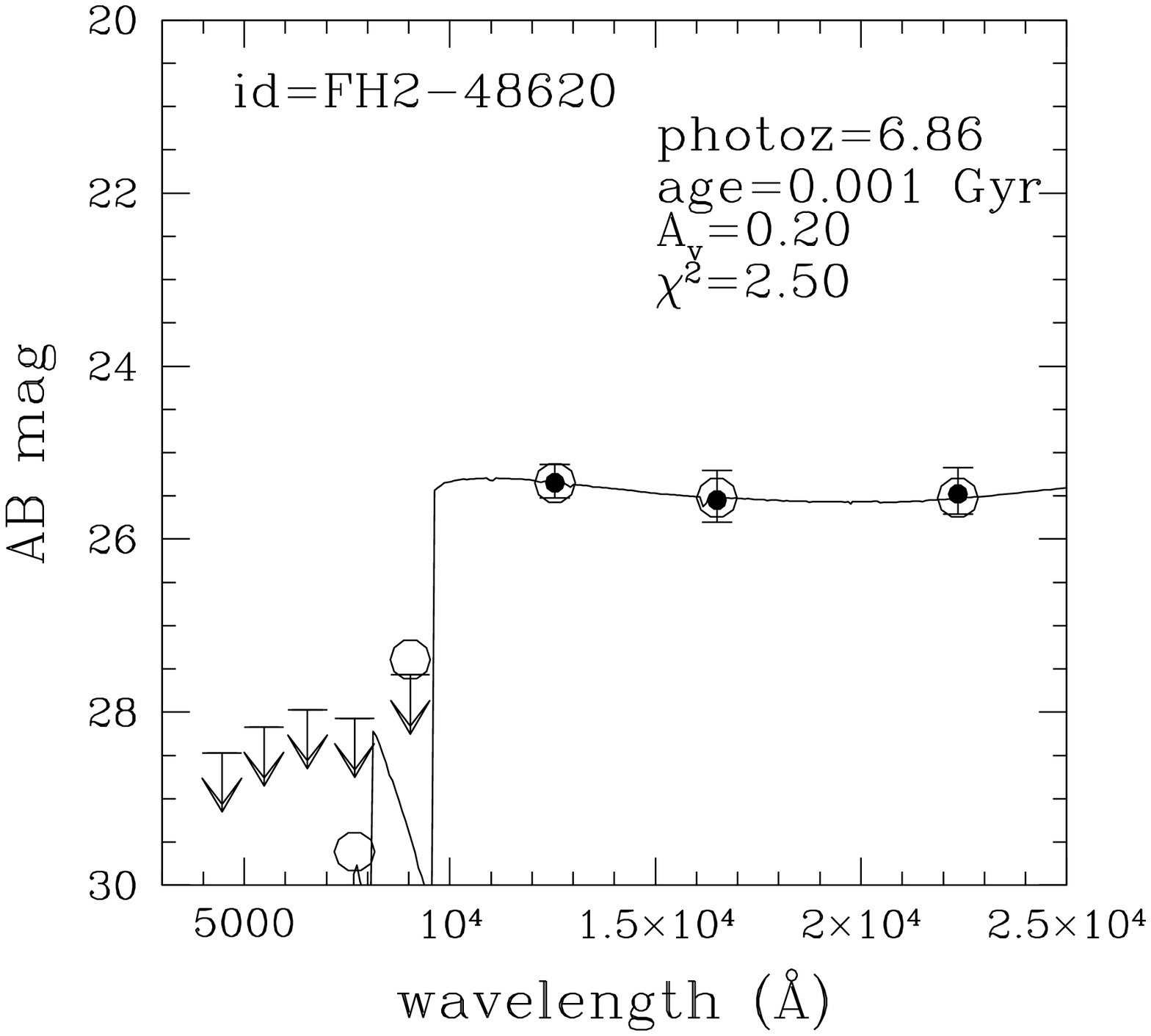}\\
\plotone{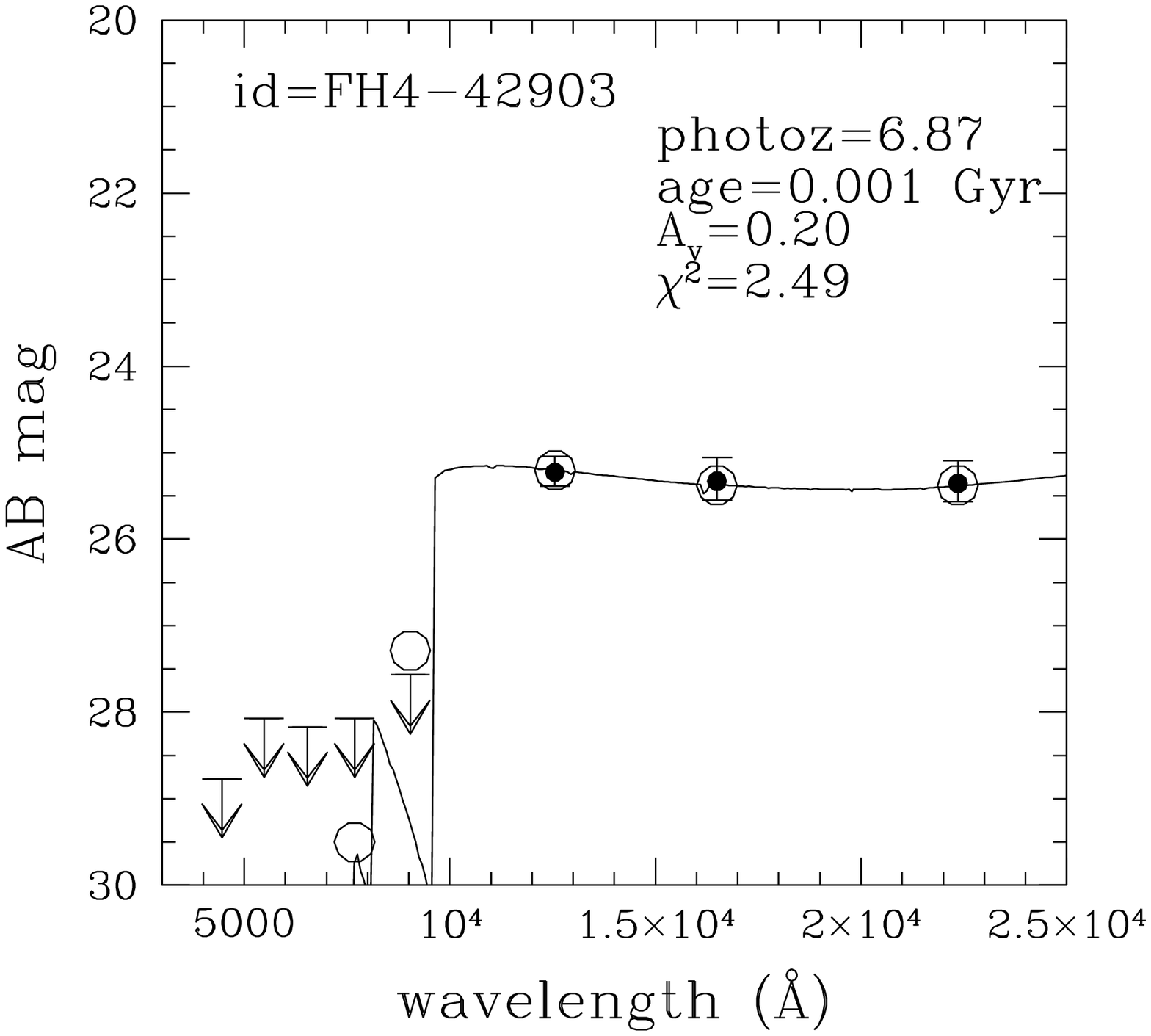}\\
\caption{Best-fit SEDs by the photometric redshift fitting. 
 In each panel, the best-fit SED (solid line) determined by 
$z_\mathrm{phot}$ fitting is plotted with the 2-$\sigma$ flux upper limits
 (arrows) in the optical bands and the fluxes in the near-infrared bands (filled circles with
 error bars), with the basic parameters derived. 
The open circles represent the convolved model fluxes, which are compared
 with the observed fluxes. The vertical axis shows the AB magnitude of
 the data points and the SEDs. Since the source 42903 has $z'$ magnitude
 (28.2) fainter than the 2-$\sigma$ flux limit, the upper limit has been
 plotted in the bottom panel, although the source is slightly visible in
 the $z'$-band image.\label{fig:photoz_fit}}
\end{figure}

We finally selected three new candidates of galaxies at $z\sim 7$ in the 
UDS field, which are listed in Table~\ref{tab:uds_cand}. 
Figure~\ref{fig:stamps} shows postage stamp images of the three sources.
The three candidates, FH2-22303, FH2-48620, and FH4-42903, 
have bright $J$ magnitudes (24.3, 25.4, 25.3; 2-arcsec aperture), 
 or rest-frame UV magnitudes of $-22.6$, $-21.8$, 
and $-21.2$, respectively. Photometric redshifts were derived
using data from all of the broad bands from $B$ to $K$ bands based on the same 
method employed in \citet{furusawaj11}.
The resultant $z_\mathrm{phot}$ are $7.05^{+0.08}_{-0.06},
6.86^{+0.13}_{-0.17}$, and $6.87^{+0.15}_{-0.17}$ for FH2-22303,
FH2-48620, and FH2-42903, respectively, supporting the high redshifts 
$z\sim 7$ of the candidate galaxies. 
The associated errors correspond to their $68\%$ confidence levels.
The best-fit spectral energy distributions (SEDs) for the three sources
are shown in Figure~\ref{fig:photoz_fit}.
The first two sources (FH2-22303 and FH2-48620) are brighter than 
all $z=7$ galaxies studied to date by spectroscopy for the Ly$\alpha$
fraction \citep{cassata11,cassata15,ono12}, if confirmed at $z=7$.
We confirmed that the number density of the combined data set
($-23<M_{UV}<-22$) of the sample by \citet{bowler14} and 
the new 3 sources $\sim (1\pm{0.4})\times 10^{-6} \rm{mag}^{-1}
\rm{Mpc}^{-3}$, is consistent with the UV LF presented by \citet{bowler14}.

\subsection{Final Photometric Sample of the $z=7$ Candidates}
We selected a total of 24 photometric sample of galaxies at
$z>6.5$. The sample consists of 19 galaxies in the
UltraVISTA field, 2 in the UDS field presented by \citet{bowler14}, 
and the 3 new galaxies in the UDS field added in this study. 

Of the 24 LBG candidates at $z=7$, 18 galaxies have very bright 
UV magnitudes of $M_{UV}<-21.75$ (14 galaxies in the UltraVISTA, 
and 4 in the UDS fields). As discussed in
Section~\ref{sec:spec}, nine galaxies are spectroscopically 
followed up in this study, and seven of these belong to the very
bright magnitude range.

\section{Follow-up Spectroscopy}\label{sec:spec}
\subsection{Observations, Data, and Analysis}
We conducted spectroscopic observations of nine sources 
of the 24 $z=7$ candidate galaxies selected in the previous
section. 
These targets were chosen to include the four high-priority sources 
in the UltraVISTA field, which were qualified as `robust' in
\citet{bowler12}, and all the five sources in the UDS field.
The `robust' targets were categorized based on results of
 $z_\mathrm{phot}$ analysis that did not accept either galactic stars or low-$z$ dusty
galaxies in \citet{bowler12}.  
A summary of the target sources is shown in
Table~\ref{tab:spec_fluxlimit}. The IDs of all the sources presented in \citet{bowler14} are taken from Table~2 of \citet{bowler14}
with the prefix `B14-', containing four sources in the UltraVISTA field
and two in the UDS field. The three sources with the prefix 'FH' are the
sources in the UDS field added in this study.
The observations were performed on the four nights of 2013 March 5 and 6 and 
2014 October 24 and 25, in multi-object spectroscopy (MOS) mode with 
Subaru/FOCAS \citep{kashik02}. 
The VPH900 grism combined with the OH58 order-sort filter was employed,
covering a wavelength range of 7,500 to 10,450 \AA\, giving a spectral resolution of 
$R\sim 1,500$ (0.74\AA\ pixel$^{-1}$) with a slit width of 0.8 arcsec. 
This combination has the highest throughput at the target wavelengths
among the 8-m class telescopes, especially at around $\sim 1\micron$,
and is therefore highly suitable for the follow-up spectroscopy of Ly$\alpha$
emission at $z=7$. Each MOS mask has a single target source at $z\sim
7$. The integrated times of 1.3 to 6 hours are devoted to each source
(Table~\ref{tab:spec_fluxlimit}), 
being split into individual exposures of 1,200 s, with a dithering 
width of 1.0 arcsec in the spatial direction between exposures. 

Data analysis was performed in a regular manner using the standard 
pipeline FOCASRED and MOSRED, implemented based on IRAF and provided by
the observatory. 
The raw spectrum data were first bias-subtracted and flatfielded by domeflat data.
The pixel areas of target spectra were extracted by applying distortion
correction based on a predefined pattern.
Then, wavelengths were calibrated based on 
night sky OH emission lines embedded in the 2D spectra, and sky
background was subtracted from the 2D spectra. 
The reduced 2D spectra of individual exposures were shifted in
the spatial direction by the dithering width and combined by taking 
median with 3-$\sigma$ clipping of outlying pixels.

Flux calibration was performed on the stacked 2D spectra 
using spectrophotometric standard stars GD153, G191B2B, and Feige110, 
with IRAF tasks `standard', `sensfunc', `fluxcalib', and `extinction'. 
With these tasks, the atmospheric extinction was corrected assuming the
atmospheric attenuation as a function of wavelength at Maunakea, which 
was measured by CFHT. 
This procedure also removed features of the flatfield as a function of
wavelength. The overall fitting error was $\sim 2\%$ rms in this
calibration, which is good enough for the following discussion.
We determined a spatial range of the source on the calibrated 2D
spectrum using the IRAF task `prows', and extracted a 1D spectrum 
 by summing fluxes within a spatial range of 6 pixels or 1.2 arcsec
centered at the flux peak.

The effects of slit-loss on the measured fluxes of each
source were corrected by estimating the missing fluxes based on the seeing
size during the spectroscopic observations and the intrinsic sizes of
the sources on the $J$-band images. The estimated slit-loss fractions
ranged from $0.23$ to $0.31$.

\subsection{Upper Limit of Ly$\alpha$ Fluxes}
\begin{figure}
\epsscale{1.}
\plotone{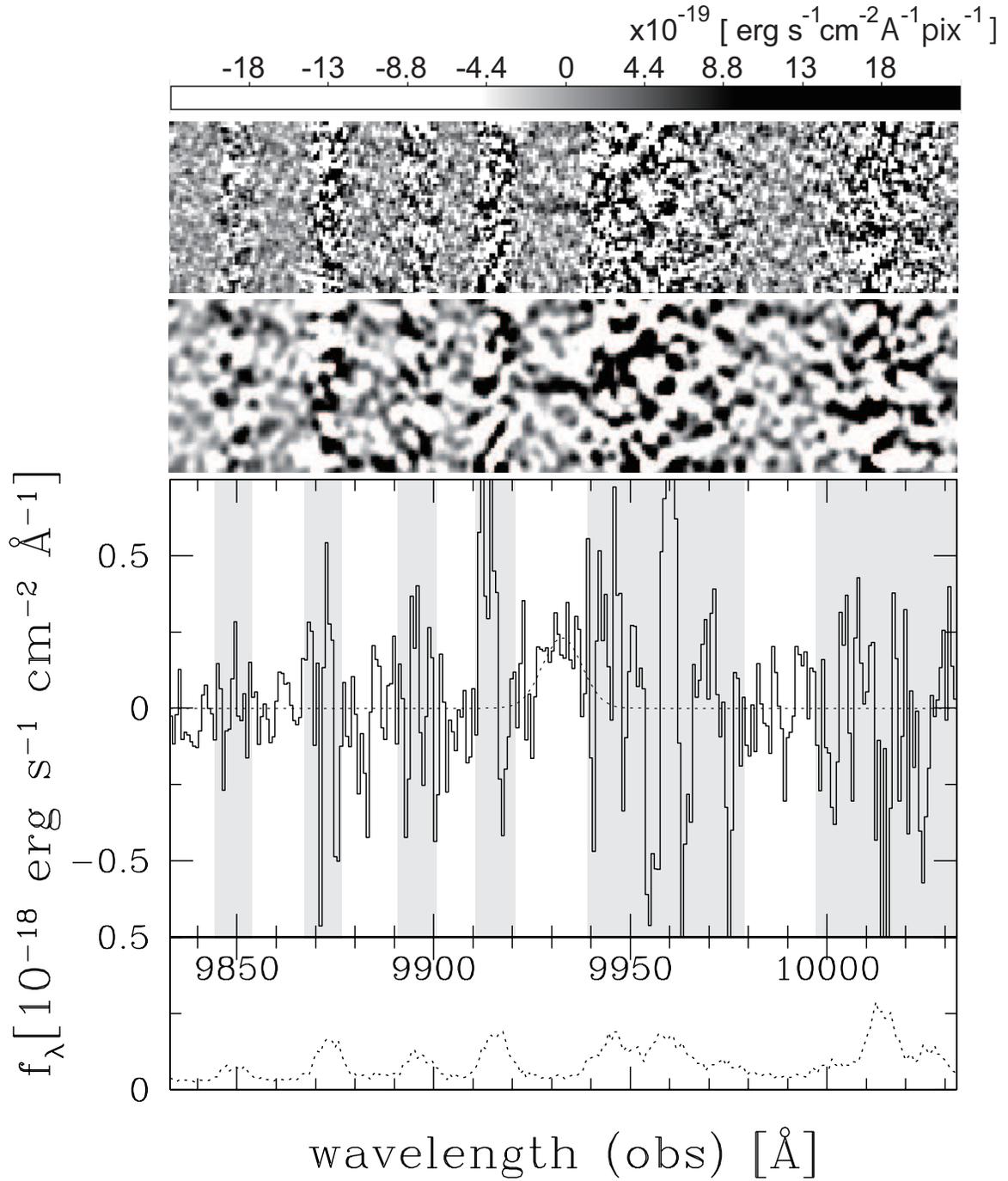}
\caption{Spectrum of the target source B14-065666 associated with 
possible Ly$\alpha$ emission in the gray scale. The top two panels
  show the 2D spectrum convolved by the Gaussian kernels with a
 $\sigma=1$ pixel (first panel) and a $\sigma=3$ pixel (second
 panel). The associated gray-scale bar indicates the flux scale of the top
 panel, and darker pixels correspond to larger intensities in the spectra.
The spatial coverage in the vertical axis of both the spectra is 12
 arcsec.
The third panel shows the 1D spectrum around the wavelength range of 
the emission line, in which the dotted line denotes the best-fit Gaussian to
 the line profile with the shaded regions overlaid on the wavelengths
 affected by the OH night sky emission lines. The 1-$\sigma$ error
 per wavelength estimated on the 2D spectrum is plotted in the bottom
 panel. All of the panels are centered on the emission line at 9,932.7\AA\ 
in the observer's frame. The S/N of the whole line profile is
 $5.5$.\label{fig:b14-065666spec}}
\end{figure}

Only a source B14-065666 displays a possible Ly$\alpha$ emission 
at $z\sim 7$, while the other eight sources showed no convincing 
emission line features to the S/N level of 3, i.e., no detection of
Ly$\alpha$. 
Figure~\ref{fig:b14-065666spec} shows the 2D and 1D spectra
of the B14-065666 at around wavelengths including the emission line
profile.
It was confirmed that the line is detected in each of 2D spectra 
generated separately for different dither positions, and spatial 
positions of the line are consistent with the dithering pattern.
Only a single line feature is detected in the source B14-065666. 
This suggests that the line is unlikely to be H$\alpha$ $\lambda
6563$\AA , as [NII]$\lambda\lambda$6548,6584\AA (S/N$\lesssim 2$) 
lines are not visible within the covered wavelength range on 
the spectrum. In addition, the red color of $z'-J\sim 2.9$ cannot be explained by ordinary galaxy SEDs.
This line is unlikely to be the [OIII], because the doublet lines 
[OIII]$\lambda\lambda$4959,5007 $\sim 95\,{\rm \AA}$ apart could 
be resolved assuming line ratio$\sim 0.4$ in the areas that are
free from the OH emission, although the S/N ($\gtrsim 2$) is not sufficiently large. 
In addition, the red $z'-J$ color cannot be explained due to 
lack of any break feature around the wavelength range.
The [OII] line in this case was located out of the wavelength
coverage and could not be detected. 
The likelihood of the [OII] is also small, as the double lines 
[OII]$\lambda\lambda$3727,3729\AA\ (5.3\AA\ apart in the observer's
frame) should be resolved, although the S/N may not be sufficiently
large assuming line ratio $\sim 1$ near an OH emission band. 
Nonetheless, the photometric redshift of
$7.04^{+0.16}_{-0.11}$\citep{bowler14} and the red $z'-J$ color of the 
 source could exclude being a lower-$z$ galaxy at $z=1.67$. 

The property of the possible Ly$\alpha$ emission was measured on 
the 1D spectrum of B14-065666. 
By profile fitting with the Gaussian function by the IRAF task `splot',
the line profile is centered at $9,932.7\,{\rm \AA}$, which corresponds to 
a redshift of 7.168 for Ly$\alpha$. 
This spectroscopic redshift agrees with the $z_\mathrm{phot}$ estimated
by \citet{bowler14}.
The measured Ly$\alpha$ flux density is $(4.4\pm 0.8)\times 10^{-18}$
erg s${}^{-1}$ cm${}^{-2}$ (S/N of 5.5), 
and the Ly$\alpha$ luminosity was $L_\mathrm{Ly\alpha}=(2.6\pm 0.5)\times
10^{42}$ erg s${}^{-1}$ (Table~\ref{tab:spec_fluxlimit}).
Here, the flux error was estimated by fluctuation of sky background 
counts over the wavelength range of the 2D spectra where the line
 profile is located, avoiding the severe OH skylines.
This luminosity is fainter than $L^*(\mathrm{Ly\alpha})$ of the Ly$\alpha$ LF 
in the redshift range by a factor of $\sim 2$ \citep{ouchi10} to $11$ \citep{matthee15}. 

As the other eight sources did not show any possible emission lines, 
we studied noise statistics of their 2D spectra to estimate the observational
upper limits of their Ly$\alpha$ fluxes.
To measure the noise fluctuation, we sampled about 500 locations on the 2D
spectra in the wavelength range corresponding to Ly$\alpha$ at $z=6.75$
to $7.25$ for five sources, including both regions free from and
affected by the OH skylines.
Each sampling area covered a rectangular area of 1 arcsec in the 
spatial direction and a width of 25\AA\ in the wavelength direction, 
as a typical Ly$\alpha$ observed line width.
For the other three sources (B14-169850, B14-035314, B14-118717), 
$z_\mathrm{phot}$ of which are below $6.75$ \citep{bowler14}, 
we chose the sampling wavelength range to cover the redshifts spanning 
$z_\mathrm{phot}\pm 0.25$. This range corresponds to the maximum
1-$\sigma\ z_\mathrm{phot}$ error of all the photometric sample 
galaxies as a conservative choice.

The histogram of the measured counts was fitted by the Gaussian function
and the 1-$\sigma$ error was estimated for each candidate source 
in the same manner as adopted in \citet{ono12}. We calculated 
the 3-$\sigma$ upper limits of the Ly$\alpha$ flux density,
which are also shown in Table~\ref{tab:spec_fluxlimit}. 

\subsection{EW Upper Limit}
We derived the upper limit of the Ly$\alpha$ flux EW for our candidate
sources based on the UV luminosities and the 3-$\sigma$ upper limit 
of the Ly$\alpha$ flux determined in the previous section. 
The EW upper limit in the observer's frame ($\mathrm{EW_{obs}}$) is 
converted to that in the rest-frame ($\mathrm{EW}_0$) on the assumption of $z=7$, except 
the source of B14-065666 ($z=7.168$). The estimated upper limits of
$\mathrm{EW}_0$ range from $1.8$ to $10.7\,{\rm \AA}$, which are shown in
Table~\ref{tab:spec_fluxlimit}.
The UV magnitudes were calculated from the $J$-band magnitudes derived 
from {\tt MAG\_AUTO} and assuming a flat spectrum of galaxies
located at $z=7$.
For the only source B14-065666 exhibiting a possible Ly$\alpha$ 
emission line, the line flux was converted to $\mathrm{EW}_0$ using a UV
magnitude of $-22.3$ as a continuum flux, assuming redshift of $7.168$.
The resultant $\mathrm{EW}_0$ is $3.7^{+1.7}_{-1.1}\,{\rm \AA}$, which 
is consistent with the tendency for UV-bright galaxies to have small
$\mathrm{EW}_0$ even in this very bright magnitude range ($<-21.75$).

\section{EW of UV Very Bright Galaxies}
\begin{figure}
\epsscale{.80}
\plotone{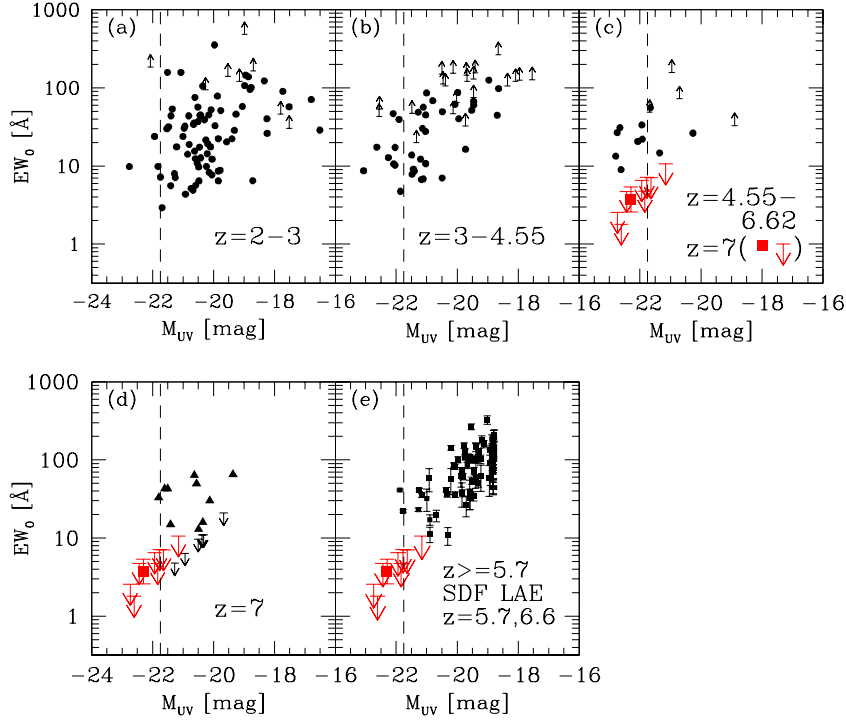}
\caption{Relation between Ly$\alpha$ $\mathrm{EW}_0$ and UV magnitudes in different redshifts. 
The 4 panels (a)-(d) show $\mathrm{EW}_0$ of Ly$\alpha$ emitting galaxies as a
 function of the absolute UV magnitude, in different redshift bins at
 $z=2-3, 3-4.55, 4.55-6.62$ and $7$. 
In panels (a)-(c) for $z<6.62$, the data points shown with filled
 circles and the lower limit represented by upward small arrows are
 taken from \citet{cassata11}.
The sample is mainly based on serendipitous detection of Ly$\alpha$
 lines in the VVDS survey \citep{lefevre13}. 
In panel (c), the data points by \citet{cassata11} are
taken for the redshift range $z=4.55-6.62$, while the upper limits to
$\mathrm{EW}_0$ in this study at $z=7$ are shown with large downward arrows 
(red), with the filled square (red) with error bars representing the
 measurement of Ly$\alpha$ emission line found in the source B14-065666
 at $z=7.168$. 
The bottom left panel (d) shows the same data points in our study at
 $z=7$ compared with those taken from \citet{ono12} (filled triangles
 and small arrows). 
In each panel, the vertical dashed line represents 
the $M_{UV}=-21.75$, which refers to the brightest magnitude for
 studying the luminosity dependence of Ly$\alpha$ $\mathrm{EW}_0$ used in
 previous studies.
Panel (e) compares our candidate galaxies with the results on 
 the LAE sample by a narrow-band survey covering a $\sim 0.25$ sq.degree field 
(small squares with error bars; \citet{kashik11}).\label{fig:ew0_z_scatter}}
\end{figure}

The relations between Ly$\alpha$ $\mathrm{EW}_0$ and UV magnitudes of
galaxies in different redshift bins are shown in
Figure~\ref{fig:ew0_z_scatter}. Our data are compared with 
previous studies, which provide the following results.

First, in panels (a)--(c) for $z<6.62$, the galaxy sample by
\citet{cassata11} includes galaxies with very bright magnitudes
($M_{UV}<-21.75$). However, only a few galaxies show 
very small $\mathrm{EW}_0$ of Ly$\alpha$ emission in this magnitude range.

Second, in panels (c) and (d), our data are shown with the upper limits 
to the $\mathrm{EW}_0$(Ly$\alpha$), which provide a new galaxy sample 
 to the very bright magnitude range at $z\sim 7$. These galaxies 
are associated with very small $\mathrm{EW}_0$ of $<10\,{\rm \AA}$. 
 Our data suggest the existence of a considerable fraction of galaxies
 with low $\mathrm{EW}_0$(Ly$\alpha$) at $z\sim 7$ at the brightest UV magnitudes
 ($<-21.75$) studied to date; only a source with $\mathrm{EW}_0>30\,{\rm
 \AA}$ at $z=7$ in this magnitude range was reported \citep{ono12} 
prior to this study.

In panel (c) $z=4.55-6.62$, the data points by \citet{cassata11} 
seem to be located at systematically higher $\mathrm{EW}_0$ than our data. 
However, their data points at $M_{UV}<-21.75$ in panel (c)
should be at $z<5.96$ based on their EW (540 to 1250\AA ) in the
observer's frame. In addition, with an independent sample based on
photometric redshifts, \citet{cassata15} indicated that $80\%$ of bright galaxies
($M<M^*$) at $z=4.5-6$ have small $\mathrm{EW}_0$ of $<30\,{\rm \AA}$,
 including galaxies with very small $\mathrm{EW}_0$ ($<10\,{\rm \AA}$).
Therefore, there should be no marked difference 
in the relation of $\mathrm{EW}_0$ and UV magnitudes at $z>6$ between 
the previous studies \citep{cassata11, cassata15} and this study.

Finally, panel (e) compares our results for the LBG candidates
with those on the narrow-band selected LAEs in a previous study 
that surveyed a $\sim 0.25$ sq.degree field \citep{kashik11}.
Our UV-selected galaxies show lower $\mathrm{EW}_0$ than the LAEs in the same
redshift range.

\section{Ly$\alpha$ Fraction of UV Very Bright Galaxies\label{sec:lya_frac}}
We aim to obtain a new constraint on the Ly$\alpha$ fraction 
for the very bright magnitude range at $z=7$, which is studied here 
for the first time.
The new result should be complementary to previous results at
fainter magnitudes to understand a luminosity dependence of the Ly$\alpha$ 
fraction.

To derive our results according to the same formalism as adopted in 
previous studies \citep{stark10, ono12, cassata15}, we use the fraction 
of galaxies with a rest-frame Ly$\alpha$ $\mathrm{EW}_0$ larger than 50\AA, 
$X_\mathrm{Ly\alpha}^{50}$.
Previous studies discussed $X_\mathrm{Ly\alpha}^{50}$ in the two
magnitude ranges (bright: $-21.75<M_{UV}<-20.25$, and faint: $-20.25<M_{UV}<-18.75$).
As seven sources of our nine sample galaxies are brighter
(Table~\ref{tab:spec_fluxlimit}) than the above bright magnitude range, 
we introduce a new magnitude range of `very bright' ($-23.0<M_{UV}<-21.75$).
We chose the threshold $\mathrm{EW}_0$ of 50\AA , as the
$X_\mathrm{Ly\alpha}^{50}$ in the bright magnitudes ($M<-21.75$) at
$z\lesssim 6$ presented by \citet{stark10} is only the previous result
that can be directly compared with our data at $z=7$.
It is essential to compare our Ly$\alpha$ fraction at $z=7$ with 
those at lower redshifts to discuss a change in the Ly$\alpha$ fraction
and progress in the reionization at these epochs.

We estimated the $X_\mathrm{Ly\alpha}^{50}$ as follows. 
First, the seven sources with UV magnitudes spanning from $-22.7$ to
$-21.8$ ($<-21.75$) were selected.
Then, weights for each source were derived by the inverse of detection 
completeness of each source as a function of brightness.
The derived weighted numbers were summed, yielding a weighted
total number of sample galaxies of $7.94$ for all seven sources.
Finally, as none of the sample galaxies show Ly$\alpha$
emission with $\mathrm{EW}_0>50\,{\rm \AA}$, we estimated upper limit of
the Ly$\alpha$ fraction.
Here, as an upper limit of the number of sources,  
we adopted a $1$-$\sigma$ upper confidence level (1.84) for the 
observed count zero based on the Poisson distribution \citep{gehrels86}.
The $X_\mathrm{Ly\alpha}^{50}$ were estimated by dividing this upper
 limit $1.84$ by the weighted total number of sample galaxies $7.94$.

In this procedure, the completeness values for the four sources in the
UltraVISTA field and one in the UDS field were taken from \citet{bowler14}. 
We conducted a Monte Carlo simulation to determine the completeness
 of the other sources in the UDS fields (Section~\ref{sec:uds_selection}). 
Each simulation was performed by adding 1,000 artificial point 
sources with known magnitudes at random positions on the $J$-band image, 
and detecting them again. The completeness was calculated as 
the fraction of recovered sources among the input sources. 
The simulations were repeated 20 times and the completeness of each
simulation was averaged to determine the effective completeness. 
We confirmed that the completeness values from \citet{bowler14} and this
study are consistent and the possible uncertainty if any should not
change the following discussion.
We did not correct for an effect of the OH skylines on
detection of the Ly$\alpha$ emission assuming the finding by
\citet{ono12}. Their simulation showed that over $90\%$ of 
simulated Ly$\alpha$ lines were recovered and the effect could be
ignored at the same wavelengths as those in this study. 
Moreover, even if B14-065666 were not a true Ly$\alpha$
detection, the following results on Ly$\alpha$ fraction would not be changed.

\begin{figure}
\epsscale{0.9}
\plotone{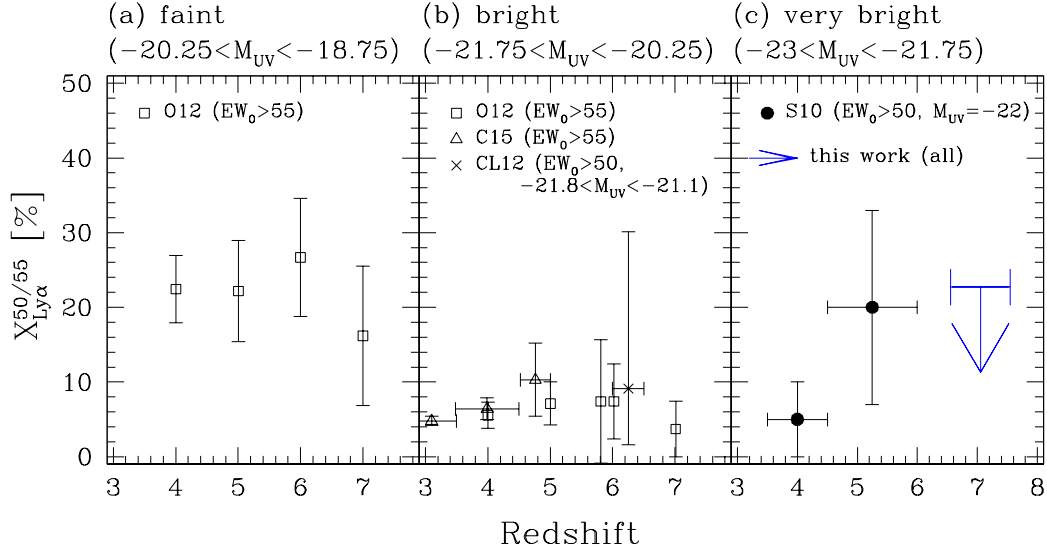}
\caption{Evolution of Ly$\alpha$ fractions of galaxies that have Ly$\alpha$ 
emission with $\mathrm{EW}_0\gtrsim 50\,{\rm \AA}$ in the three magnitude ranges. 
From left to right, the Ly$\alpha$ fractions of UV selected 
galaxies in various studies are shown over the redshifts $z=3$ to $7$ 
in different magnitude bins of (a) faint ($-20.25<M_{UV}<-18.75$), 
(b) bright ($-21.75<M_{UV}<-20.25$), and (c) very bright ($-23<M_{UV}<-21.75$).
In the (a) faint and (b) bright magnitudes, we show the compilation of
 results from previous studies. The result ($\mathrm{EW}_0\gtrsim 55\,{\rm
 \AA}$) in the (a) faint magnitudes is derived from \citet{ono12}: open
 squares, based on \citet{fontana10,pentericci11,schenker12,stark11}. 
In the (b) bright magnitudes, the data points are taken from the three studies, 
\citet{ono12}: open squares, \citet{curtis-lake12}: cross, and \citet{cassata15}: 
open triangles, where the data by \citet{ono12} are composites of
 their own data and those of
 \citet{dow07,stanway07,fontana10,vanzella11,pentericci11,stark11,schenker12}.
The Ly$\alpha$ fractions by \citet{ono12} and \citet{cassata15} are
 derived from their results for $\mathrm{EW}_0>55\,{\rm \AA}$. 
We obtain the data point for \citet{curtis-lake12} at $z=6-6.5$ using
 their complete sample of 11 galaxies at $-21.8<M_{UV}<-21.1$ that
 showed Ly$\alpha$ $\mathrm{EW}_0>50\,{\rm \AA}$, 
with error bars estimated by simple Poisson uncertainty.
In the right panel for the (c) brightest magnitude range ($-23<M_{UV}<-21.75$), 
we plot the upper limit of the Ly$\alpha$ fraction ($\mathrm{EW}_0>50\,{\rm \AA}$) in this work 
at $z=7$ determined with the seven galaxies (solid blue arrow), compared
 with those of \citet{stark10}: filled circles. 
Our results are estimated for the sources at $-23<M_{UV}<-21.75$, while
 those by \citet{stark10} at the lower redshifts of $z=3.5-6$ are 
 derived from the same magnitude range at $M_{UV}\sim -22$.
The different thresholding EW of $50$ or $55\,{\rm \AA}$ among the studies 
is due to the difference in available data in the literature. This
 small difference would not change the trends of the Ly$\alpha$ fraction
 in different magnitude ranges.  
\label{fig:xlya50_z}}
\end{figure}

Figure~\ref{fig:xlya50_z}\ shows changes in the Ly$\alpha$ fraction
($\mathrm{EW}_0\gtrsim 50\,{\rm \AA}$) over $z=3-7$ in the three
different UV magnitude ranges: (a) faint ($-20.25<M_{UV}<-18.75$), (b)
bright ($-21.75<M_{UV}<-20.25$), 
and (c) very bright ($-23<M_{UV}<-21.75$).
Our data provide upper limits to the $X_\mathrm{Ly\alpha}^{50}$ at $z=7$
in the very bright range for the first time. 
The following findings are obtained.

First, in the (c) very bright magnitude range, our data point at $z=7$
possibly implies that the Ly$\alpha$ fraction at $z=7$ remains at a
similar fraction or lower than that at $z=6$, although the data points
have large errors. In the same panel, the data points derived from \citet{stark10}
may support a rapid increase in $X_\mathrm{Ly\alpha}^{50}$ from $z=3.5$
to $z=6$. 
\citet{ono12} suggested that the Ly$\alpha$ fraction shows a significant
decline at $z=7$ from $z=6$, which is also seen in panels (a) and (b) at
the fainter magnitudes.  Our results are consistent with this
finding even in the very bright magnitude range.

Second, in comparison between panels (b) and (c),  
our upper limit may indicate a weak trend of 
the Ly$\alpha$ fraction in the very bright magnitude range
($-23<M_{UV}<-21.75$) at $z=7$ consistently being at 
a similar level to that at fainter magnitudes.
The possibility of a slightly higher Ly$\alpha$ fraction at $z=7$ in the
(c) very bright range than in the (b) bright range also cannot be 
excluded. 
This result seems to extend the relation at $z\sim 4.5-6$ \citep{stark10} 
where the Ly$\alpha$ fraction at very bright magnitudes ($M_{UV}\sim
-22$) is at the same level or slightly higher than that of 
fainter magnitudes ($-21.75<M_{UV}<-20.75$). 
A similar trend was suggested by \citet{curtis-lake12} in their bright
magnitude range ($L>2L^*$) at $z\sim 6$. They suggested that 
the Ly$\alpha$ fraction for the bright sample may be significantly
higher (by $\sim 20\%$ or a factor of $\sim 2$) than that of the fainter
sample \citep{stark11}.
Our finding may present some hints of a different trend from 
 that suggested by \citet{ono12},  
where the Ly$\alpha$ fraction for the faint ((a) $-20.25<M_{UV}<-18.75$) sample is
 consistently higher than that of the bright ((b) $-21.75<M_{UV}<-20.25$) sample at $z=4-7$. 

Finally, our upper limit may imply that the Ly$\alpha$ fraction 
in the (c) very bright magnitudes at $z=7$ is comparable to or lower
than that in the (a) faint magnitude range at the same redshifts.

\section{Discussion}
\subsection{Ly$\alpha$ Fraction}
The results in the previous section could provide 
only weak implications for the Ly$\alpha$ fraction of bright 
galaxies due to the current large errors. Nevertheless, 
the trend seen in the Ly$\alpha$ 
fraction may possibly imply the existence of physical mechanisms 
that cause the LBGs with large luminosities to have an Ly$\alpha$ 
escape fraction similar to or higher than that of fainter LBGs ($M_{UV}>-21.75$).
For example, \citet{bowler14} discussed a possible excess at the 
bright end of the UV LF at $z=7$ by extrapolation of the conventional 
Schechter function. 
The similar excess of the Ly$\alpha$ LFs has also been studied by
\citet{matthee15}.
The excess in the very bright galaxies would require consideration 
of some special physical processes of galaxy formation, such as 
lack of an efficient mass-quenching mechanism \citep{peng10} in this
epoch. 
\citet{bongiorno16} implied that the feedback from 
a central active galactic nucleus (AGN) of galaxies, where star formation 
rate is suppressed by outflows from a luminous AGN, could be a
mass-quenching mechanism. If the AGN feedback is weak in the bright galaxies at $z=7$,
destruction of the H {\sc ii}
regions associated with star formation would be reduced, which supports
higher luminosities of Ly$\alpha$ emission.
At the lower redshifts, where the AGN feedback is thought to work 
more efficiently, star formation in bright galaxies is 
more suppressed than in faint galaxies. The bright galaxies
would have higher metallicity in the gas, and possibly more 
dust than in the counterparts in $z\gtrsim 7$; hence, 
the Ly$\alpha$ emission would be less efficient at the lower
redshifts.

In addition, the outflow from a galaxy, if any, may help an increase 
in the Ly$\alpha$ escape fraction, due to scattering of the Ly$\alpha$
lines, e.g., \citet{dijkstra14b}.
Moreover, the very bright LBGs may be located in massive dark matter 
halos where the reionization progresses in an earlier epoch. This trend may
lead to a higher transmission of the IGM than in less massive halos. 
Our result seems to be consistent with this trend.

It is difficult to clearly interpret the observation that 
the Ly$\alpha$ fraction in the very bright range at $z=4.5-6$ is 
still higher than that in the bright range at the same redshifts.  
This could be in part accounted for by magnification 
of the sample galaxies due to gravitational lensing by their foreground 
galaxies \citep{vanderburg10}, although the result cannot be fully explained.

We note that the above discussion depends on the uncertainty of
 measurements of the Ly$\alpha$ fraction at each redshift.
As the error bars for the very bright sample galaxies at both lower
 redshifts reported by \citet{stark10} and at $z=7$ examined in this study are 
still large, we cannot draw strong conclusions regarding the trend based
 only on the existing data, and further spectroscopic samples are
 necessary to obtain more reliable constraints.

\subsection{Cosmic Neutral Hydrogen Fraction}
Now that we have derived the Ly$\alpha$ fraction in the unprecedentedly 
bright UV magnitude range, it is interesting to derive a new 
constraint on the neutral H {\sc i} fraction ($\xHI$) in the 
cosmic reionization epoch. 
The cumulative probability distribution function (PDF) 
of Ly$\alpha$ $\mathrm{EW}_0$, which is the probability of galaxies having a 
rest-frame $\mathrm{EW}_0$ larger than a given value, has been used to discuss 
the neutral H {\sc i} fraction with faint spectroscopic samples of
galaxies ($-20.25<M_{UV}<-18.75$) at $z=7$ in previous studies
\citep{pentericci11, ono12}.
\citet{ono12} reported that the neutral H {\sc i} fraction may be 0.6 to 0.9 by 
comparing their EW-PDF with the model of \citet{dijkstra11}, which is 
consistent with those suggested by \citet{schenker12} and \citet{pentericci11}.
\citet{caruana14} obtained $\xHI\sim 0.5$ at $z=7$
with 22 relatively faint $z$-dropout galaxies ($-21.1<M_{UV}<-18.0$),
followed up by FORS2 of the Keck Telescope. The slight difference in
resultant neutral H {\sc i} fraction from that of \citet{ono12} may have
been due to difference in the magnitude ranges studied.

\begin{figure}
\epsscale{.55}
\plotone{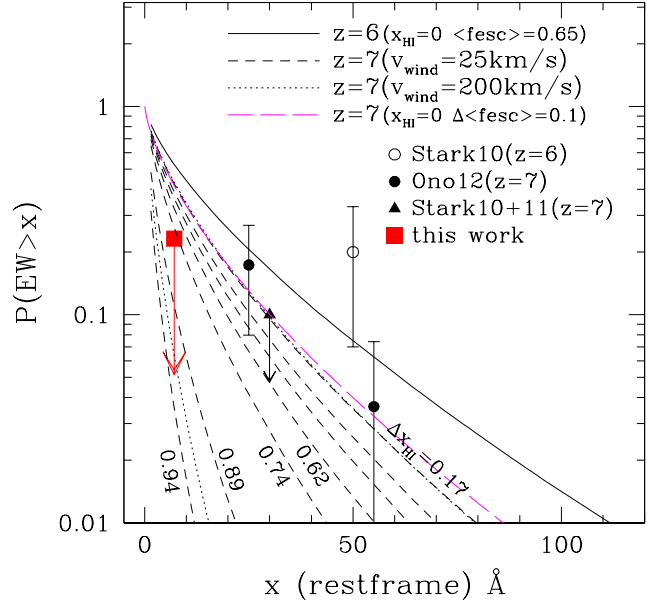}
\caption{The cumulative probability distribution function of $\mathrm{EW}_0$ (Ly$\alpha$) (EW-PDF). 
The upper limit derived in this study is shown with 
the red square symbol with downward arrow at $\mathrm{EW}_0=7.1\,{\rm \AA}$.
Overplotted here are the data points from a combined result of 
\citet{stark10,stark11} at $\mathrm{EW}_0=30\,{\rm \AA}$ for the magnitudes
 $-20.5<M_{UV}<-19.5$ which are derived from \citet{dijkstra11} (filled
 triangle), and those by \citet{ono12} at $\mathrm{EW}_0=25, 55\,{\rm \AA}$ in
 the magnitude range $-21.75<M_{UV}<-20.25$ (filled circles). 
The model prediction of the EW-PDF presented by \citet{dijkstra14} is 
compared to the observational results.
The solid curve located above a series of curves is the reference model
 EW-PDF determined at $z=6$ as an epoch after the cosmic reionization
 has almost completed ($\xHI\sim 0$). 
In the reference model, the IGM transmission is estimate assuming a
 fixed escape fraction of ionizing photons (0.65), and shell wind velocity
 of 25 km s$^{-1}$ with H {\sc i} column density of $10^{20}$cm$^{-1}$. 
The data point at $z\sim 6$ by \citet{stark10} (open circle)
 is compared, and is consistent with reference EW-PDF within 
the error bars.
We estimate the models of EW-PDFs for various neutral
H {\sc i} fractions at $z=7$ (dashed curves; the seven models of $\Delta \xHI
 = 0.17, 0.40, 0.51, 0.62, 0.74, 0.89$, and $0.94$) relative to that of 
the reference epoch $z=6$. The estimation is performed by rescaling the 
reference EW-PDF at $z=6$ by the PDFs of the IGM transmission for the
 seven $\xHI$ provided by \citet{dijkstra11, dijkstra14}, on the
 assumption that all changes in the  EW-PDF can be attributed to the change in $\xHI$, with a fixed
 escape fraction of ionizing photons.
We also plot the dotted curves predicted with a different shell wind model 
(200 km s$^{-1}$) only for the two $\xHI =0.17$ and $0.94$, 
suggesting only small differences between the two wind velocities across
 the $\xHI$ range studied. 
The long-dashed curve (magenta) shows, for reference, the EW-PDF 
with the increased escape fraction by 0.1 with no change in the $\xHI$,
from $z=6$ to $7$.\label{fig:pdf_ew0}}

\end{figure}

Figure~\ref{fig:pdf_ew0} shows the resultant EW-PDF of our new
spectroscopic sample in the very bright UV magnitude range
($-23<M_{UV}<-21.75$), 
compared with the results reported in the literature at the same redshift $z=7$.
We estimated the upper limit of EW-PDF at $\mathrm{EW}_0=7.1\, {\rm \AA}$ as follows.
Of our seven spectroscopic sample galaxies in the very bright
magnitude range (i.e., without B14-118717 and FH4-42903), the largest value of the 
observational upper limit of the $\mathrm{EW}_0$ was $7.1\,{\rm \AA}$ (FH2-48620;
Table~\ref{tab:spec_fluxlimit}). 
Hence, we assume that the Ly$\alpha$ emission with $\mathrm{EW}_0>7.1\,{\rm
\AA}$, if any, would be detected in the seven spectroscopic data, but
none of the seven sources shows such an emission line. Therefore, 
the EW-PDF upper limit was derived by dividing the
$1$-$\sigma$ upper limit $1.84$ by the weighted total number of sample 
galaxies $7.94$, to EW-PDF($\mathrm{EW}_0>7.1\,{\rm \AA}$)$\sim 0.23$ (filled
square with downward arrow).
Here, we assume the non-detection of EW to 3-$\sigma$ 
significance based on the noise statistics averaged over the studied 
wavelength range, which includes both areas with and without OH 
emission. This assumption may be slightly optimistic for this small EW
range, as there is a chance that possible lines, if any, would be 
  located in the relatively noisier OH emission bands. Nevertheless, the
  essence of the following discussion is not changed by the uncertainty.

Although we assume $z=7$ for all the sample galaxies 
except $z=7.168$ of B14-065666, our discussions in this study 
would also not be changed by making use of $z_\mathrm{phot}$ in estimation of
$\mathrm{EW}_0$. The typical difference in the upper limits determined 
by $z=7$ and $z_\mathrm{phot}$ is sufficiently small ($\Delta EW\sim 0.1\,{\rm
\AA}$).

We compare our data with the model prediction of EW-PDF presented by 
\citet{dijkstra14}. 
Their model was generated by computing PDF of a fraction 
of Ly$\alpha$ photons transmitted through the IGM, combining models of 
galactic shell-wind outflows with large-scale semi-numeric simulations of
reionization.
They determined the reference model EW-PDF at $z=6$ ($\xHI\sim 0$) 
to coincide with the median Ly$\alpha$ fraction ($\mathrm{EW}_0>75\,{\rm \AA}$)
at $z=6$ by \citet{stark10}.
The data point at $z\sim 6$ by \citet{stark10} in the very bright
magnitude bin at $M_{UV}=-22$ (open circle) is also plotted in
Figure~\ref{fig:pdf_ew0}, which is still consistent with the reference
EW-PDF within the error bars.
We estimated the EW-PDF models for the various neutral H {\sc i} 
fractions at $z=7$ by rescaling the reference EW-PDF (see caption of
Figure~\ref{fig:pdf_ew0}).
It is noted that the model of \citet{dijkstra14} 
was calculated for the UV magnitude range $-21.75<M_{UV}<-20.25$,
which is fainter than that of our data, due to the
dark matter halo mass range considered in their calculation. 
Nevertheless, inclusion of more massive halos, which is naively 
interpreted as the inclusion of UV-brighter galaxies, should 
produce little if any change in the result (Mark Dijkstra, private
communication).
Hence, we simply compare their model with our data.

Our upper limit would not contradict the model curves of
$\xHI\gtrsim 0.7$. Although this is a speculative comparison 
due to the large error and the small sample size,  
our data may possibly imply $\xHI$ around $0.7$ to $0.9$ 
with the assumed escape fraction of $\sim 0.65$. This $\xHI$ is 
consistent with the report of \citet{ono12} in the fainter magnitude range.
This result in the new very bright magnitude range may give 
further support to the findings discussed in previous studies that 
the reionization of the universe progresses rapidly from $z=7$ to $=6$. 
We examined the same comparison with the models at fainter
magnitudes ($-20.75<M_{UV}<-18.75$) using the data points in \citet{ono12}, 
and found an inferred $\xHI$ of
$\lesssim 0.7$. The combination of the results in the very bright and 
faint magnitudes may favor $\xHI\sim 0.7$ on the
assumption of the escape fraction $\sim 0.65$.

Based on our data together with the previous results of \citet{ono12}
and \citet{stark10} at the same redshift, the possibility of lower
$\xHI\sim 0.2$ cannot be excluded. 
As shown by the long-dashed curve (magenta), evolution of the escape
fraction of ionizing photons from galaxies may contribute to the net
change in EW-PDF. The possible increase in escape fraction may
significantly reduce the Ly$\alpha$ fraction under the same IGM
transmission,  which would lower the $\xHI$ required to explain the
observed EW-PDF at $z=7$.

Moreover, \citet{dijkstra14} mentioned the effect on their model 
of the possible existence of LBGs with Ly$\alpha$ absorption
($\mathrm{EW}<0$) at high redshifts $z>6$, which was reported by
\citet{shapley03} for $z=3$ LBGs.
They suggested that the EW-PDF model would be shifted by 
$\Delta EW_0\sim -25\,{\rm \AA}$ by considering LBGs with
$\mathrm{EW}<0$. 
Such a shift may explain the Ly$\alpha$ fraction ($\mathrm{EW}_0=25\, {\rm \AA}$,
$M_{UV}<-20.25$) at $z=8$ of $0.07-0.08$ presented by \citet{treu13}. 
If this is the case at $z=7$ in this study, the 
possible $\xHI$ range would be further decreased to $0.2$ or lower.
However, if we rescale the $z=6$ reference EW-PDF model to exactly match 
the data point of \citet{stark10} (open circle), 
even higher H {\sc i} fractions than the $\xHI\sim 0.7$ would be favored.
Together with the above discussion, the lack of any strong Ly$\alpha$,
such as $>25\,{\rm \AA}$, in our spectroscopic sample may imply a hint
of the unlikely very low $\xHI$ at $z=7$.

Thus, further investigations to gain a better understanding of the physical 
properties and evolution of galaxies at high redshifts are 
necessary to provide stringent constraints on $\xHI$. 
In addition, the number of spectroscopic samples of galaxies is still limited, 
and therefore future efforts to increase the size of the sample from low to high
luminosity ranges, for reliably determining the Ly$\alpha$ fraction will
be indispensable.

\section{Conclusions}
We conducted a search for UV-bright LBGs in two 
legacy survey fields, UltraVISTA and UKIDSS-UDS, which cover an area of 
1.65 sq.degree ($J=25.3-25.5$). 
Very deep imaging observations on the two survey fields have been undertaken 
in the $z'$ band ($z'\sim 26.5$; 5$\sigma$, 2-arcsec aperture) 
with Subaru/Suprime-Cam.

We performed selection of $z=7$ candidates in the UDS field from the
multi-waveband catalogs ($J=25.5$; 5$\sigma$) with the updated $z'$-band data, based
on the red color in $z'-J$, combined with relatively mild to red color in $J-K$,
 and dropout in all optical bands. 
In the UDS field, we chose three candidates of possible UV-bright galaxies at $z=7$.
The 19 candidate sources in the UltraVISTA field 
and two sources in the UDS field at $z>6.5$ presented by \citet{bowler14} 
were combined with the three new sources in the UDS field, yielding 24
targets in total.
The spectroscopic observations were obtained on nine sources of the
$z=7$ candidates with Subaru/FOCAS. 
Only a single source B14-065666 shows possible Ly$\alpha$ emission 
at $z=7.168$, while the other eight sources show no emission line features. 
By measurements of noise fluctuations on the 2D spectra, 
the observational upper limits of 
the 3-$\sigma$ upper limits of the Ly$\alpha$ fluxes and $\mathrm{EW}_0$ 
were estimated based on these eight sources.
The upper limits of Ly$\alpha$ $\mathrm{EW}_0$ span $1.8$ to $10.7\,{\rm \AA}$ for
these sources.
The only source showing a possible Ly$\alpha$ emission, 
B14-065666, has an estimated $\mathrm{EW}_0$ of $3.7^{+1.7}_{-1.1}\,{\rm \AA}$ 
with a UV magnitude of $-22.3$. This result supports a tendency 
for UV-bright galaxies to likely have small $\mathrm{EW}_0$ even 
at very bright magnitudes ($<-21.75$). 

The upper limits of Ly$\alpha$ EW at $z=7$ were compared with those of
previous studies at $z=2-6$. 
Our new data provide clear upper limits to the $\mathrm{EW}_0$(Ly$\alpha$) 
at $z\sim 7$ in the very bright UV magnitude $<-21.75$, implying that
 a considerable fraction of galaxies with very bright magnitudes
 ($<-21.75$) have low EW at $z\sim 7$.

Based on the upper limits of the Ly$\alpha$ EW, 
the 1-$\sigma$ upper limits of Ly$\alpha$ fraction with 
thresholding $\mathrm{EW}_0=50\,{\rm \AA}$ ($X_\mathrm{Ly\alpha}^{50}$) at $z=7$
was derived for the brightest magnitude range studied to date ($-23.0<M_{UV}<-21.75$).
While the Ly$\alpha$ fraction may support a rapid increase from $z=3.5$
to $z=6$ even in this bright magnitude range, 
it may possibly imply leveling off, remaining at 
the same level or below at $z=7$. This result may support 
the findings at fainter magnitudes reported by \citet{ono12}.
Our data may also indicate a weak trend whereby the Ly$\alpha$ fraction at
$-23<M_{UV}<-21.75$ may be similar to that of the fainter magnitudes 
at $z>5$, possibly up to $z=7$. A slightly higher fraction cannot be
ruled out. With a large uncertainty, this result may witness the earlier
progression of reionization in more massive dark matter halos, as well
as a possible physical mechanism for providing a higher Ly$\alpha$
escape fraction in the very UV-bright galaxies at redshift 7. 

Finally, we discussed the Ly$\alpha$ EW-PDF derived from 
the upper limits of the Ly$\alpha$ fraction at $z=7$, to provide a new 
constraint on the neutral H {\sc i} fraction ($\xHI$). 
Our resultant EW-PDF at $\mathrm{EW}_0=7.1\,{\rm \AA}$ at the 
very bright magnitudes was combined with previous studies at 
fainter magnitudes by \citet{stark10,stark11} and \citet{ono12}.
We performed a speculative comparison of the observed EW-PDF 
with the model of \citet{dijkstra14}.
Although the constraint is not strong given the large error 
bars, we derived the neutral H {\sc i} fraction 
of $\xHI =0.7-0.9$ favored by our data, leaving open the 
possibility of lower $\xHI$ depending on evolution of the physical
 properties of galaxies, such as the escape fraction.
The same comparison with EW-PDF for the fainter counterparts
($-20.75<M_{UV}<-18.75$) may support $\xHI \lesssim 0.7$, which is 
consistent with or below the value for the brighter
galaxies.
However, further compilation of spectroscopic observations 
is necessary to determine more reliable constraints on the Ly$\alpha$
fraction and reionization at $z\gtrsim 7$ based on a firm understanding 
of the evolution of galaxies. 

\acknowledgments
We thank the anonymous referee for helpful comments that 
have improved the manuscript.
We are grateful to Rebecca Bowler for her valuable contribution to 
the proposals for spectroscopic observations. We express our gratitude
to Yutaka Ihara, Yuko Ideue, Nana Morimoto for their contributions to
imaging observations, data reduction, and data evaluation, and Kimihiko Nakajima for data evaluation. 
We are grateful to Takashi Hattori for supporting observations and 
data reduction. We thank Omar Almaini for preparing the UDS data, and 
Mark Dijkstra for providing his latest theoretical model of the
Ly$\alpha$ EW-PDF. 
The Subaru imaging observations were assisted by Yoshiyuki Inoue, Misae
 Kitamura, Kohki Konishi, Mariko Kubo, Yu Niino, and Takahiro Ohno.
We thank Satoshi Miyazaki and Suprime-Cam team for commissioning 
Suprime-Cam FDCCDs, which has realized this work. This work is based 
in part on data collected at Subaru Telescope, 
which is operated by the National Astronomical Observatory of Japan.
UltraVISTA is based on data products from observations made with ESO
Telescopes at the La Silla Paranal Observatory under ESO programme 
ID 179.A-2005 and on data products produced by TERAPIX and the Cambridge 
Astronomy Survey Unit on behalf of the UltraVISTA consortium.
Data analysis were in part carried out on common use data analysis computer 
system at the Astronomy Data Center, ADC, of the National Astronomical 
Observatory of Japan. 
UKIDSS uses the UKIRT/WFCAM (Casali et al. 2007) and a 
photometric system described in Hewett et al. (2006), and the calibration is described in Hodgkin et al. (2009). The pipeline processing and science archive are described in Irwin et al. (2016, in prep) and Hambly et al. (2008). 
JSD acknowledges the support of the European Research Council via the award 
of an Advanced Grant.
This work is partly supported by JSPS KAKENHI Grant Number 23740159 (HF)
and 15H03645 (NK).

{\it Facilities:} 
\facility{Subaru (Suprime-Cam, FOCAS)},
\facility{UKIRT (WFCAM)}, 
\facility{ESO:VISTA (VIRCAM)}.

\clearpage

\clearpage

\clearpage

\clearpage

\clearpage

\clearpage

\clearpage

\clearpage

\begin{deluxetable}{lrrr}
\tabletypesize{\scriptsize}
\tablecaption{Summary of $z'$-band Imaging Data\label{tab:zdata}}
\tablewidth{0pt}
\tablehead{
\colhead{Field Name} & \colhead{Total Exposure} & \colhead{Limiting Magnitude (5$\sigma$, $\phi 2''$)} & \colhead{PSF FWHM\tablenotemark{a}}\\
\colhead{} & \colhead{(min)} & \colhead{(mag)} & \colhead{(arcsec)}
}
\startdata
UltraVISTA1 & 1173 &  26.63 & 0.76 \\
UltraVISTA2 & 1096 &  26.46 & 0.84 \\
UltraVISTA3 & 1123 &  26.47 & 0.70 \\
UltraVISTA4 & 1250 &  26.75 & 0.82 \\
UDS1 & 701 & 26.30 & 0.75 \\     
UDS2 & 970 & 26.57 & 0.81 \\
UDS3 & 703 & 26.60 & 0.78 \\
UDS4 & 928 & 26.58 & 0.81 \\
\enddata
\tablenotetext{a}{The FWHMs of PSF sizes in each image before geometric transformation.}
\end{deluxetable}

\clearpage

\begin{deluxetable}{lccrrrrrrrr}
\tabletypesize{\scriptsize}
\tablecaption{Summary of New $z=7$ Candidate Objects in the UDS Field\label{tab:uds_cand}}
\tablewidth{0pt}
\tablehead{
\colhead{Object} & \colhead{RA(J2000)} & \colhead{Dec(J2000)} & \colhead{$B$} & \colhead{$V$} & \colhead{$R_c$} &
 \colhead{$i'$} & \colhead{$z'$} & \colhead{$J$} & \colhead{$H$} &
 \colhead{$K$} \\
}
\startdata
FH2-22303 & 02:16:25.092 & $-$04:57:38.50 & $>$28.7\tablenotemark{a,b} &
 $>$28.2 & $>$28.0 & $>$28.0 & $>$27.6 & $24.3\pm 0.1$  &
 $24.2\pm 0.1$ & $24.4\pm 0.1$ \\ 
FH2-48620 & 02:17:39.083 & $-$04:42:48.71 & $>$28.8 & $>$28.1 & $>$28.2 & $>$28.1 & $>$27.6 & $25.4\pm 0.2$ &
 $25.6\pm 0.3$ & $25.5\pm 0.2$ \\ 
FH4-42903 & 02:17:26.306 & $-$05:10:16.27 & $>$28.5 & $>$28.2 & $>$28.0 & $>$28.1 & $>$27.6 & $25.3\pm
 0.2$ & $25.4\pm 0.2$ & $25.4\pm 0.2$ \\ 
\enddata
\tablenotetext{a}{The listed magnitudes are measured within 2-arcsec apertures.}
\tablenotetext{b}{The upper limits in the optical bands are 
the 2-$\sigma$ limiting magnitudes, which are derived from the 
public SXDS data \citep{furusawa08} for the $B, V, R_c, i'$ bands, and
 converted from Table~\ref{tab:zdata} for the $z'$ band.}
\end{deluxetable}

\clearpage

\begin{deluxetable}{lrrrrr}
\tabletypesize{\scriptsize}
\tablecaption{Summary of Spectroscopic Sample of Galaxies\label{tab:spec_fluxlimit}}
\tablewidth{0pt}
\tablehead{
\colhead{Object} & \colhead{Total Exposure} & \colhead{Flux Limit (3$\sigma$)} & \colhead{$L_\mathrm{Ly\alpha}$ Limit (3$\sigma$)} & \colhead{$M_{UV}$} & \colhead{$\mathrm{EW}_0^\mathrm{Ly\alpha}$ Limit (3$\sigma$)} \\
\colhead{} & \colhead{(s)} & \colhead{(erg s$^{-1}$ cm$^{-2}$)} & \colhead{(erg s$^{-1}$)} & \colhead{(mag)} & \colhead{(\AA )}
}
\startdata
B14-169850 & 4800 & $6.9\times 10^{-18}$ & $3.9\times 10^{42}$ &  $-22.4\pm{0.1}$ & 4.7 \\
B14-065666 & 14400 & $(4.4\pm 0.8)\times 10^{-18}$\tablenotemark{a}
 & $(2.6\pm 0.5)\times 10^{42}$\tablenotemark{a} &
 $-22.3\pm{0.2}$ & $3.7^{+1.7}_{-1.1}$\tablenotemark{b} \\
B14-304416 & 7200 & $4.9\times 10^{-18}$ & $2.8\times 10^{42}$ & $-22.7\pm{0.1}$ & 2.6 \\
B14-238225 & 6000 & $6.0\times 10^{-18}$ & $3.4\times 10^{42}$ & $-21.9\pm{0.2}$ & 6.5 \\
B14-035314 & 7200 & $4.0\times 10^{-18}$ & $2.3\times 10^{42}$ & $-21.8\pm{0.2}$ & 4.7 \\
B14-118717 & 6740 & $4.9\times 10^{-18}$ & $2.8\times 10^{42}$ & $-21.6\pm{0.2}$ & 7.1 \\
FH2-22303 & 19200 & $3.1\times 10^{-18}$ & $1.8\times 10^{42}$ & $-22.6\pm{0.1}$ & 1.8 \\
FH2-48620 & 4800 & $5.6\times 10^{-18}$ & $3.2\times 10^{42}$ & $-21.8\pm{0.2}$ & 7.1 \\
FH4-42903 & 7200 & $4.8\times 10^{-18}$ & $2.7\times 10^{42}$ & $-21.2\pm{0.2}$ & 10.7 \\
\enddata

\tablenotetext{a}{Measured values of the line flux rather than the upper
 limits. The flux error is estimated by a fluctuation of sky background counts for the wavelength range of the 2D spectra where the line profile is located.}
\tablenotetext{b}{$\mathrm{EW}_0$ is estimated for a measured flux on the possible emission line at $z=7.168$.}
\end{deluxetable}

\end{document}